\documentclass[preprint,12pt]{elsarticle}

\usepackage{graphicx}
\usepackage{amssymb}

\usepackage{lineno}
\usepackage{textcomp}

\usepackage{amsmath,amsfonts,amssymb}
\usepackage{graphicx}
\usepackage[colorlinks=true, allcolors=blue]{hyperref}

\usepackage{pdflscape}

\renewcommand{\epsilon}{\varepsilon}

\newcommand{\hess}{H.E.S.S.}
\newcommand{\nectar}{NECTAr}
\newcommand{\nectarchip}{NECTAr chip}
\newcommand{\hessI}{H.E.S.S.~I}
\newcommand{\hessIu}{H.E.S.S.~I Upgrade}

\newcommand{\zmq}{{\O}MQ}

\let\seriesfb\bfseries\def\bfseries{\boldmath\seriesfb}
\let\seriesdm\mdseries\def\mdseries{\unboldmath\seriesdm}

\newcommand{\fref}[1]{Fig.~\ref{#1}}

\newcommand{\sref}[1]{Sect.~\ref{#1}}

\journal{Astroparticle Physics}

\begin{document}

\begin{frontmatter}


\title{A \nectar-based upgrade for the Cherenkov cameras\\ of the H.E.S.S. 12-meter telescopes}



\author[label2]{T.~Ashton}
\author[label8,label10]{M.~Backes}
\author[label3]{A.~Balzer}
\author[label3,label1]{D.~Berge}
\author[label5]{J.~Bolmont}
\author[label1]{S.~Bonnefoy}
\author[label4]{F.~Brun}
\author[label4]{T.~Chaminade}
\author[label4]{E.~Delagnes}
\author[label6]{G.~Fontaine}
\author[label1]{M.~F\"u{\ss}ling}
\author[label1]{G.~Giavitto\corref{cor1}}
\ead{gianluca.giavitto@desy.de}
\author[label6]{B.~Giebels}
\author[label4]{J.-F.~Glicenstein}
\author[label1]{T.~Gr\"aber}
\author[label2,label7]{J.A.~Hinton}
\author[label9]{A.~Jahnke}
\author[label1]{S.~Klepser\corref{cor1}}
\ead{stefan.klepser@desy.de}
\author[label1]{M.~Kossatz}
\author[label1]{A.~Kretzschmann}
\author[label1,label4]{V.~Lefranc}
\author[label1]{H.~Leich}
\author[label5]{J.-P.~Lenain}
\author[label1]{H.~L\"udecke}
\author[label1]{I.~Lypova}
\author[label6]{P.~Manigot}
\author[label7]{V.~Marandon}
\author[label4]{E.~Moulin}
\author[label1]{T.~Murach}
\author[label6]{M.~de~Naurois}
\author[label5]{P.~Nayman}
\author[label1]{S.~Ohm}
\author[label1]{M.~Penno}
\author[label2]{D.~Ross}
\author[label3]{D.~Salek}
\author[label1]{M.~Schade}
\author[label7]{T.~Schwab}
\author[label8]{K.~Shiningayamwe}
\author[label1]{C.~Stegmann}
\author[label1]{C.~Steppa}
\author[label5]{J.-P.~Tavernet}
\author[label2]{J.~Thornhill}
\author[label5]{F.~Toussenel}
\author[label5]{P.~Vincent}

\address[label1]{DESY, D-15738 Zeuthen, Germany}
\address[label2]{Department of Physics and Astronomy, The University of Leicester, University Road, Leicester, LE1 7RH, United Kingdom}
\address[label3]{GRAPPA, Anton Pannekoek Institute for Astronomy, University of Amsterdam,  Science Park 904, 1098 XH Amsterdam, The Netherlands}
\address[label4]{IRFU, CEA, Universit\'e Paris-Saclay, F-91191 Gif-Sur-Yvette Cedex, France}
\address[label5]{Sorbonne Universit\'es, Universit\'e Paris Diderot, Sorbonne Paris Cit\'e, CNRS/IN2P3, Laboratoire de Physique Nucl\'eaire et de Hautes Energies, LPNHE, 4 Place Jussieu, F-75252, Paris, France}
\address[label6]{Laboratoire Leprince-Ringuet, Ecole Polytechnique, CNRS/IN2P3, F-91128 Palaiseau, France}
\address[label7]{Max-Planck-Institut f\"ur Kernphysik, P.O. Box 103980, D 69029 Heidelberg, Germany}
\address[label8]{University of Namibia, Department of Physics, Private Bag 13301, Windhoek, Namibia}
\address[label10]{Centre for Space Research, North-West University, Potchefstroom 2520, South Africa}
\address[label9]{JA consulting, St Michael Park 23, Avis, Windhoek, Namibia}
\cortext[cor1]{Corresponding authors.}
\begin{abstract}
The High Energy Stereoscopic System (\hess) is one of the three arrays of
imaging atmospheric Cherenkov telescopes (IACTs) currently in operation. It
is composed of four 12-meter telescopes and a 28-meter one, and is
sensitive to gamma rays in the energy range $\sim30$~GeV -- 100~TeV. The
cameras of the 12-m telescopes recently underwent a substantial upgrade,
with the goal of improving their performance and robustness. The upgrade
involved replacing all camera components except for the photomultiplier
tubes (PMTs). This meant developing new hardware for the trigger, readout,
power, cooling and mechanical systems, and new software for camera control
and data acquisition. Several novel technologies were employed in the
cameras: the readout is built around the new \nectar\ digitizer chip,
developed for the next generation of IACTs; the camera electronics is
fully controlled and read out via Ethernet using a combination of FPGA and
embedded ARM computers; the software uses modern libraries such as Apache
Thrift, \zmq\ and Protocol buffers. This work describes in detail the
design and the performance of the upgraded cameras.  
\end{abstract}

\begin{keyword} Gamma-ray astronomy \sep Cherenkov camera \sep High-energy
instrumentation upgrade \sep PMT Cameras \sep \nectar \sep H.E.S.S.



\end{keyword}

\end{frontmatter}


\section{Introduction}

The first Cherenkov telescopes of the \hess \ array were the four 12-meter
diameter CT1--4, built and commissioned between 2002 and 2004 at the
\hess\ site in the Khomas highlands in Namibia (see e.g. \cite{hess}). CT1--4
are also known as the "\hessI\ array". A fifth, 28-meter diameter telescope was
built in 2012 in the centre of the square \hessI\ array. The main goal of this
new telescope, called CT5, was lowering the minimum gamma-ray energy threshold
of \hess \ from $\sim 100$~GeV down to $\sim30$~GeV. To reach that goal, CT5
has a very large mirror area (614~m$^2$), photosensors with higher quantum
efficiency and a camera \citep{ct5_camera,sam} with a much lower dead-time than
the original CT1--4 ones. CT5 can trigger on low energy air showers with a rate
of $\sim3$~kHz, about ten times the event rate of the older, smaller CT1--4.

An important reason to upgrade the 14-year-old CT1--4 cameras \cite{oldcameras}
was to enable the CT1--4 array to trigger at a lower threshold, resulting in
more events being recorded stereoscopically with CT5. This could not be achieved
with the original cameras because of their rather large readout dead-time of
$\sim 450$~{\textmu}s per event: lowering their trigger threshold by e.g. 30\%
would have increased the fraction of events lost due to dead-time to $\sim15$\%.

An equally important reason to upgrade the old CT1--4 cameras was to prevent the
inevitable increase of failures due to the ageing of the electronics,
connectors and other critical parts that had been exposed for 14~years to the
harsh conditions of the Namibian site. Furthermore, many electronic components
had become obsolete and could not be procured anymore, making the cameras
increasingly difficult to maintain.

This work is structured as follows: general description of design and
architecture (\autoref{sec:arch}); tests performed on individual components and
on the integrated system (\autoref{sec:test}); calibration procedures employed
for commissioning and deployment (\autoref{sec:cal}); performance achieved in
the field (\autoref{sec:perf}); conclusions (\autoref{sec:conclusions}).

%

\begin{figure}
  \centering
  \includegraphics[width=0.49\textwidth]{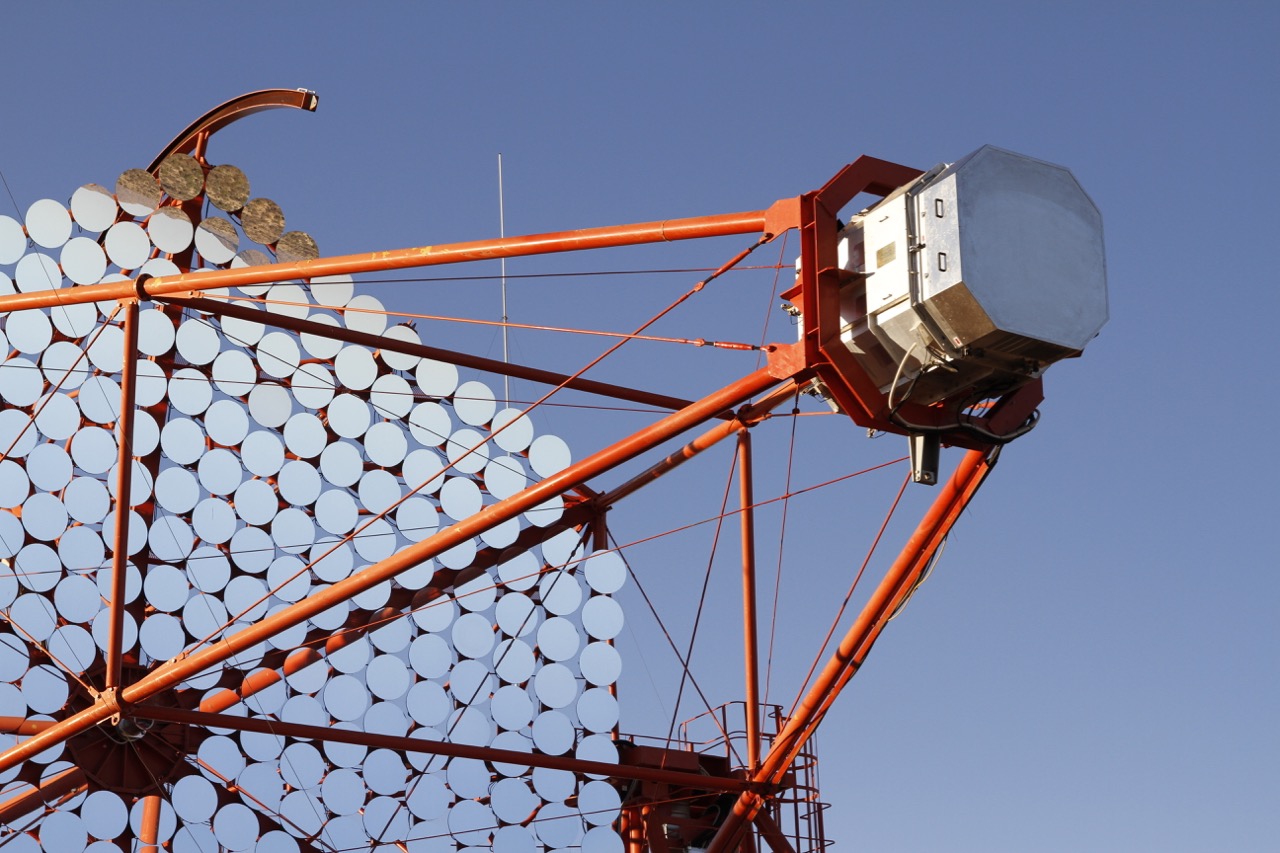}
  \includegraphics[width=0.49\textwidth]{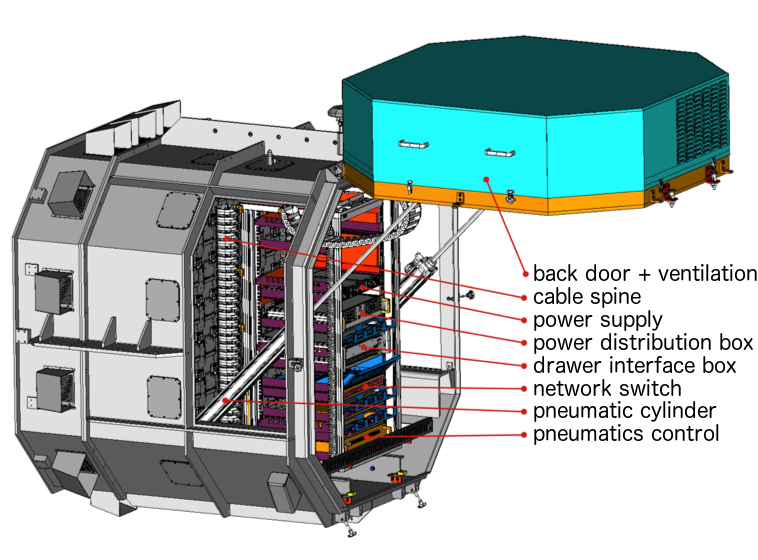}
  \caption{Left: A picture of the first upgraded \hessI\ camera, mounted on CT1.
  Right: Rear 3D view of the of the camera. The backplane rack is visible, the
  ventilation is contained inside the back door (in light blue). The mechanical
  structure of the camera was built at the LLR laboratory.}
   \label{fig:overview}
\end{figure}

\section{Architecture of the new cameras}
\label{sec:arch}

\begin{landscape}
\begin{figure}[p]
  \includegraphics[height=\textwidth]{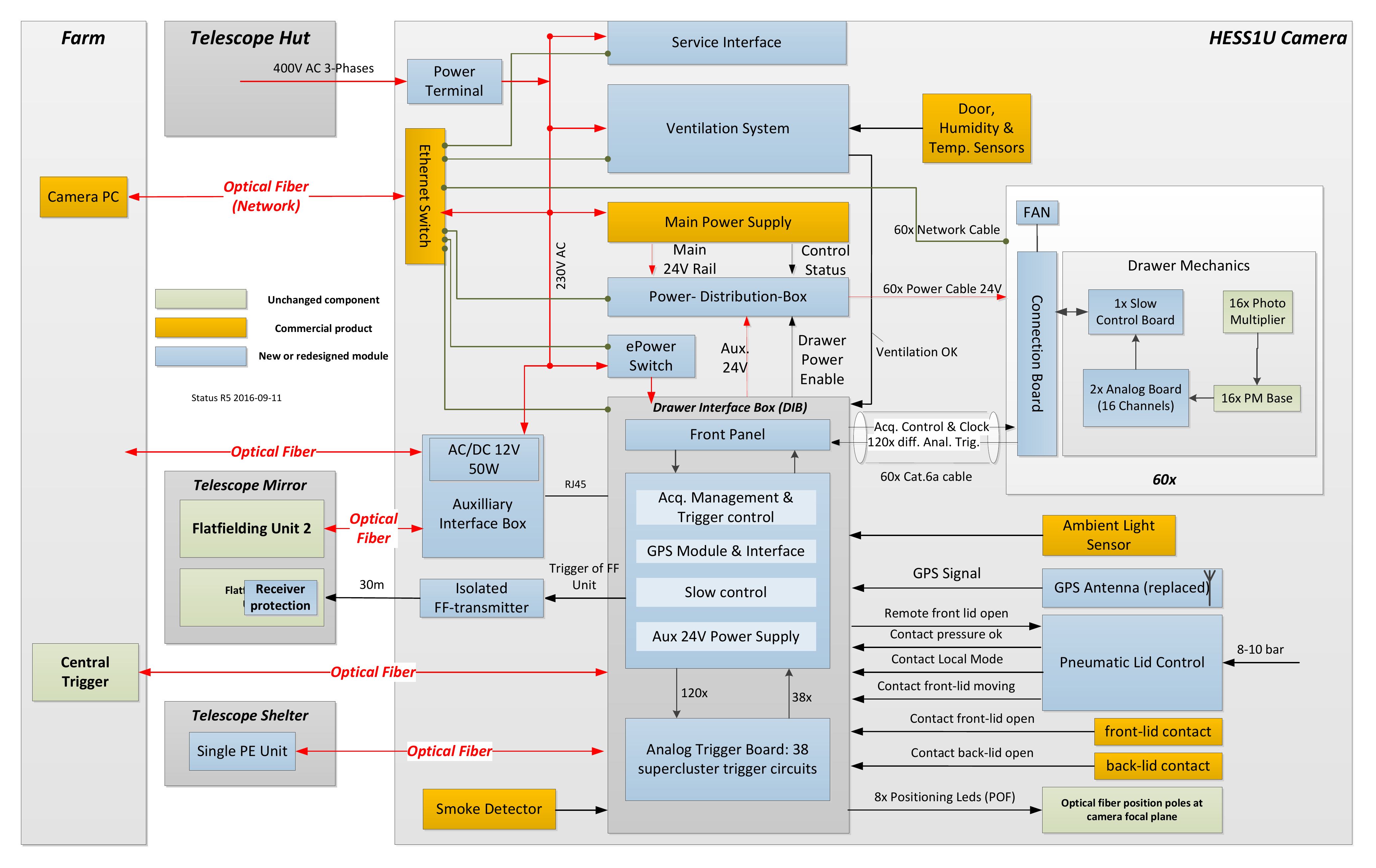}
  \caption{This diagram illustrates how the various mechanical and electronic
    subsystems of the camera interact. Original, custom-made, and commercial
    off-the-shelf components are marked in green, blue and orange,
    respectively. Red lines represent the power distribution and the arrows its
    direction; if labeled ``Optical Fiber'', they represent bidirectional
    optical fiber links. Green, circle-terminated lines represent copper
    Ethernet links. Black lines and arrows represent electrical signals.
    Physical locations of the subsystems are marked with light grey boxes:
    ``HESS1U Camera`` is the camera body, the ``Telescope Shelter'' is the
    camera daytime parking shelter; the ``Telescope Hut'' is a service
    container tied to the telescope structure, the ``Farm'' is the server room
    inside the array control building.}
  \label{fig:architecture}
\end{figure}
\end{landscape}

Upgrading the \hessI \ cameras meant replacing or refurbishing essentially every
component inside them. Only the photomultiplier tubes (PMTs) and their high
voltage power supplies (HV bases) were kept, due to their cost and relative
robustness. This can also be seen in the schematic diagram of the camera
subsystems (\fref{fig:architecture}). When possible, commercial
off-the-shelf (COTS) solutions were employed. A shared design feature of all
custom electronic subsystems developed for the cameras is the usage of an FPGA
coupled to a single-board computer, controlled via Ethernet.

Most of the development, production and testing of the cameras has been done at
the DESY site in Zeuthen. A picture of one of the upgraded cameras on the
telescope can be seen in \fref{fig:overview}, left. 

\subsection{Front-end electronics}

\begin{figure}
  \centering
  \includegraphics[width=0.8\textwidth]{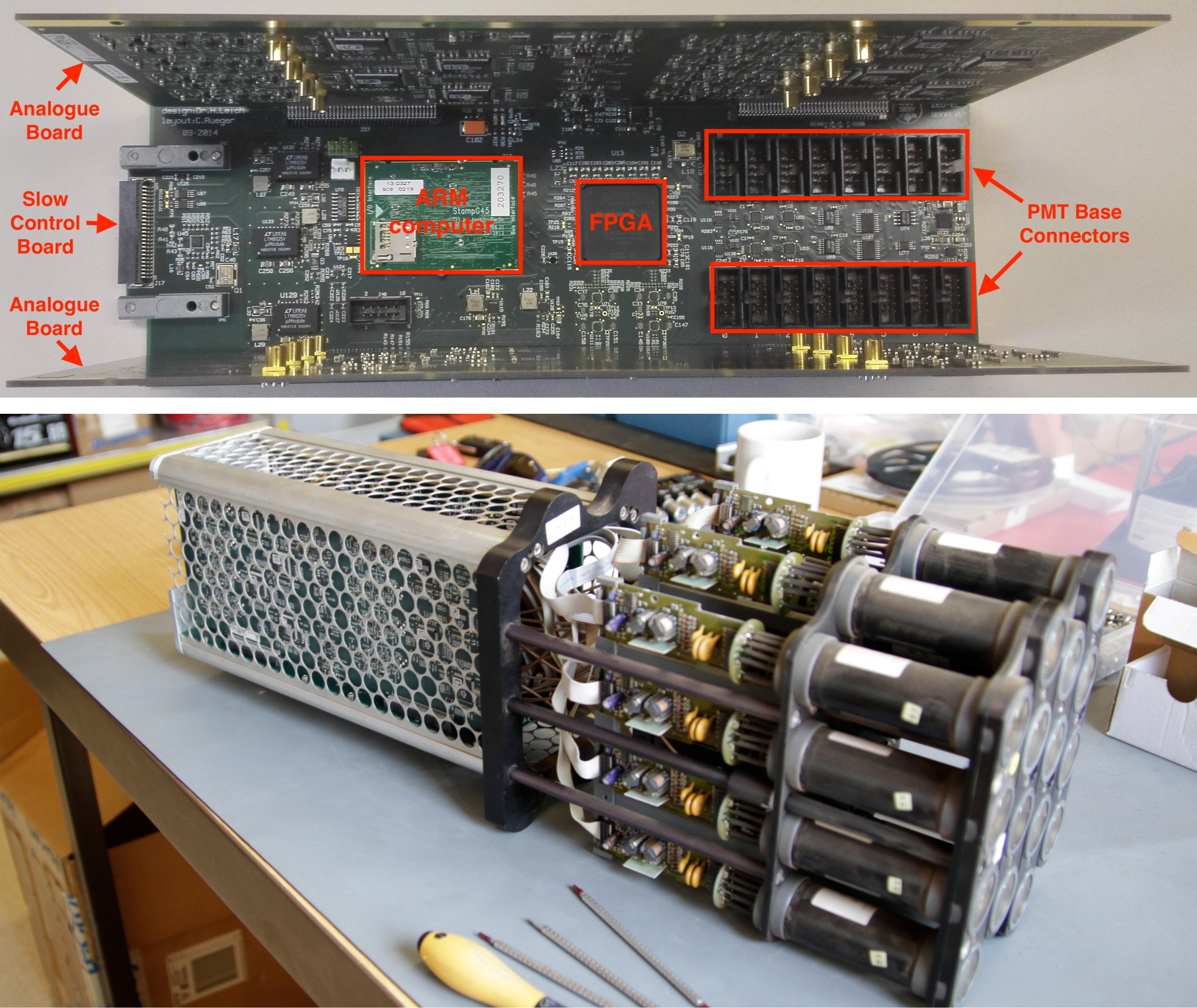}
  \caption{Top: Annotated inside view of a partially assembled drawer. 
  Bottom: A fully assembled drawer.} \label{fig:drawer}
\end{figure}

Cherenkov light from particle showers in the atmosphere is detected and
digitized in the front-end of the camera. The light sensors are 960~PMTs,
organized into 60 modules, called ``drawers''
(\fref{fig:drawer}). The drawers are arranged in a $9\times8$ rectangular
matrix, with each corner of the matrix devoid of 3 drawers. A drawer consists
of 16 PMTs, two 8-channel analogue boards, and a slow control
board (see \fref{fig:drawer}, top).  The analogue boards host the electronics
components responsible for the amplification, discrimination and digitization
of the PMT signals (see \fref{fig:nectar}, left); the slow control board hosts
an FPGA (Altera Cyclone IV) controlling the whole drawer, an ARM9-based
single-board computer (TaskIt Stamp9G45), the power regulators and the sockets
for the PMT HV bases. The drawers are supported by a mechanical structure which
separates the front from the back-end of the camera. Each drawer is connected
to a connection board secured on the back-end side of that structure, hosting
sockets for network, trigger and power (see \fref{fig:backplane}, right).

\subsubsection{Analogue boards}

The analogue signal from one PMT is sent to the analogue board via a 15~cm long
coaxial cable. The PMTs produce negative polarity, single-ended voltage pulses
of 2--3~ns duration (FWHM) with an amplitude varying from 1~mV to a few~V,
depending on the number of photons detected. Upon reaching the analogue board,
the PMT signals are AC coupled, pre-amplified by a factor 9.8, split into three branches and
further amplified by low noise single-ended to differential amplifiers, which
also invert their polarity. 
\\
Two of the branches are routed to the two inputs of
the \nectar\ readout chip, for sampling and digitization. Their overall amplification
factors are 15.1 (high gain, HG) and 0.68 (low gain, LG). The \nectarchip\
inputs have a nominal range of 2~V, so high gain signals are clipped to 3.3~V, the most convenient voltage
present on the board within the \nectarchip\ tolerance range, to avoid affecting the low gain. An adjustable
constant common-mode offset of about 0.2~V is added to the electrical signal to
keep it within the input range even in the case of undershoot (this corresponds
to a pedestal offset of around 420~ADC counts). 
\\
The signal in the third branch is amplified by a factor 45 and sent to a
high-speed comparator, whose digital output is directly routed to the FPGA on
the slow control board. This signal is referred to as the level 0 (L0) trigger
signal.

\subsubsection{Readout}

\begin{figure}
  \centering
  \includegraphics[width=0.37\textwidth]{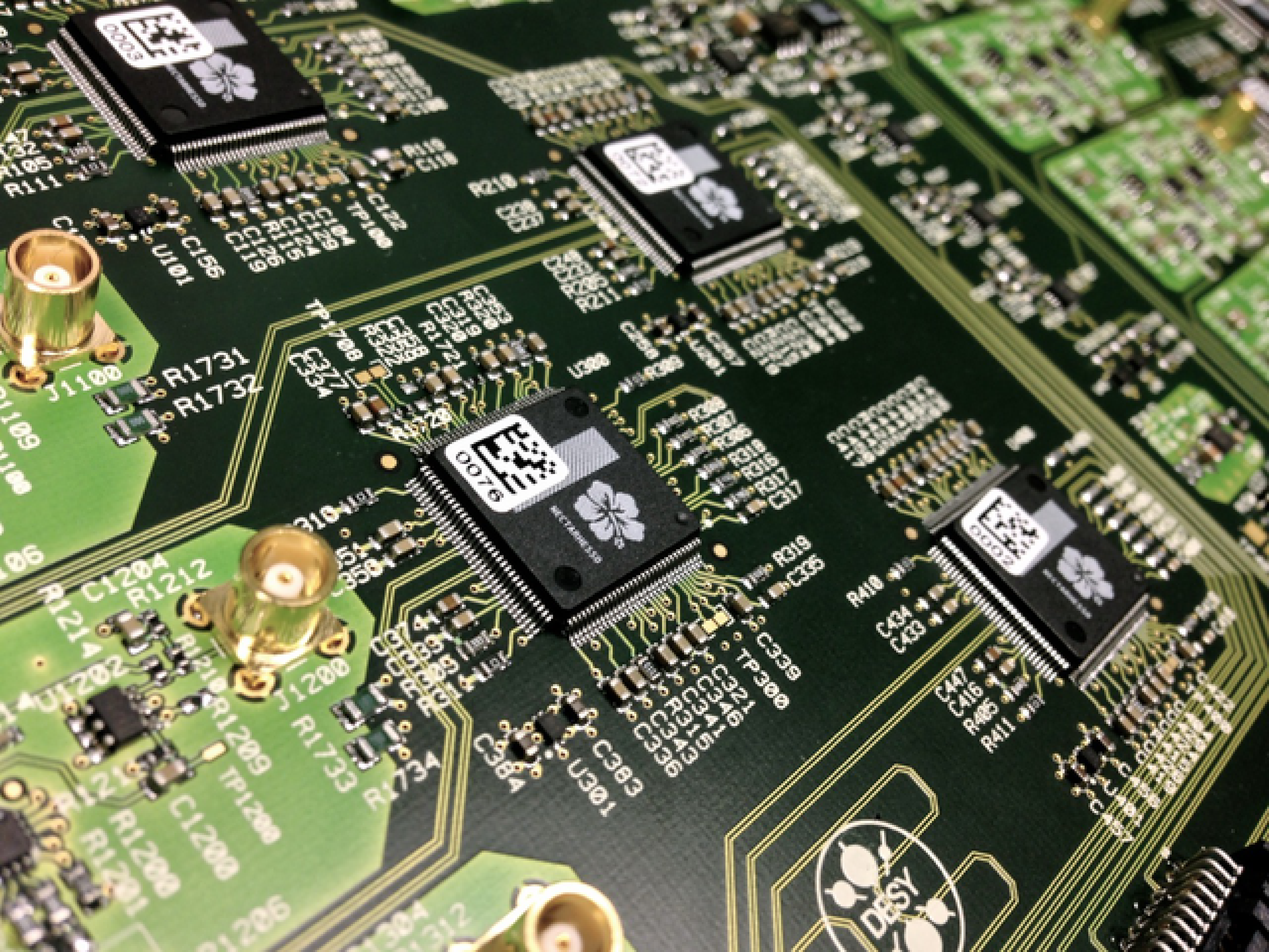}
  \includegraphics[width=0.61\textwidth]{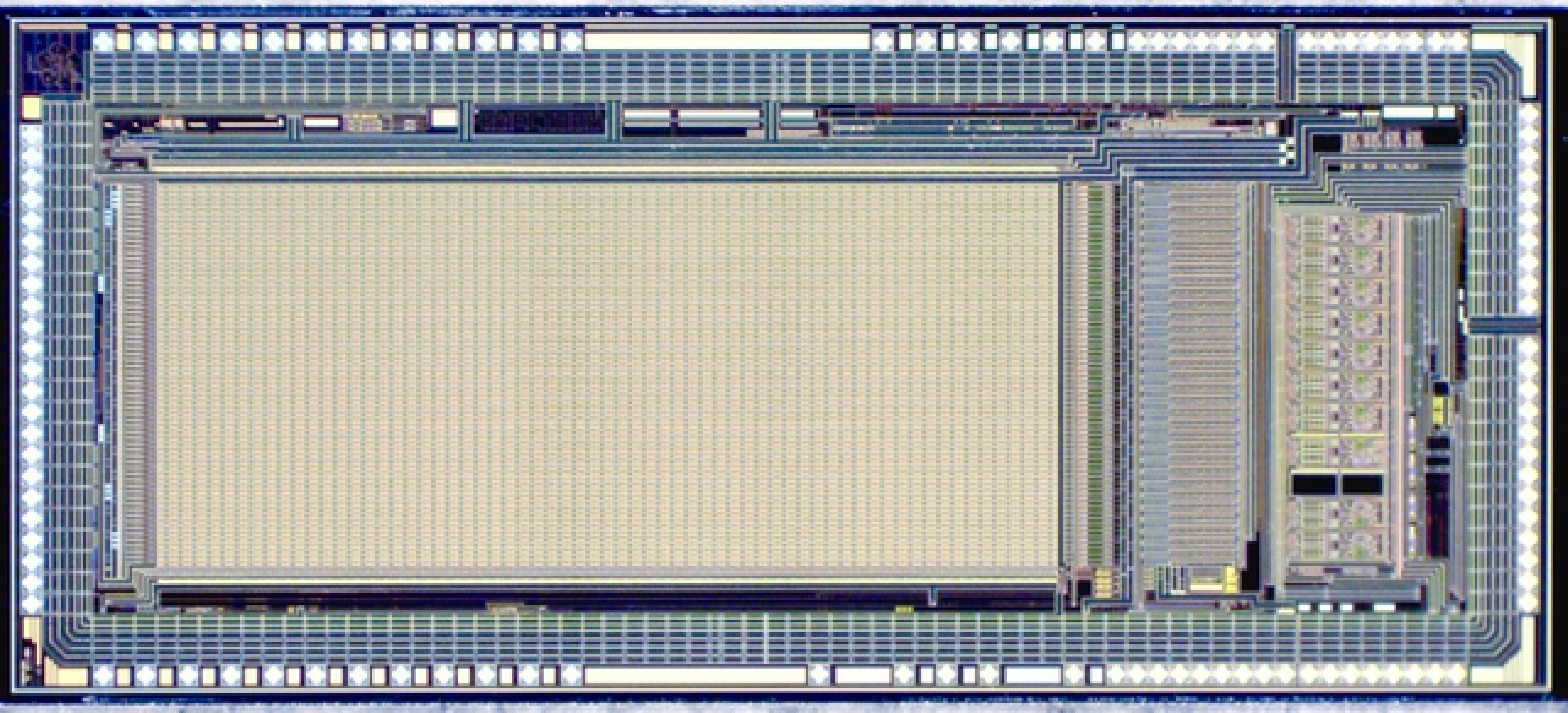}
    \caption{Left: Part of the analogue board showing the analogue
    amplification stages (light green) and the \nectar\ chips (black with a nectar
    logo). Right: Microphotography of the \nectarchip.}
  \label{fig:nectar}
\end{figure}

Most of the performance improvements of the upgraded cameras are due to its readout
electronics, based on the \nectar\ analogue memory chip, designed at CEA/IRFU
\cite{nectar} (\fref{fig:nectar}, right).

The \nectarchip\ has two channels (one per gain), each equipped with a switched
capacitor array of 1024 cells, acting as an analogue ring memory buffer. There
are two modes of operation: writing and reading. During the writing phase, the
input amplitude is stored sequentially on the array capacitors, with a switching
frequency of 1~GHz. The writing process is circular over the whole array, so the
charge stored in the cells is overwritten every 1024~ns by the new input. A
trigger signal stops the writing and initiates the reading: the charges in the
capacitor cells of a small region of interest (ROI) are read out and digitized
by the on-chip 12-bit 21~Msamples/s ADC. The digital data is then transmitted
to an FPGA by means of a serializer. For regular observations the ROI is
currently set to 16~cells, and the data in the ROI is summed by the FPGA, and
sent to the camera server as one integrated charge value per pixel and per
gain. The choice of ROI length and simple summing charge integrator is
inherited from the old cameras for compatibility with the existing \hess\
analysis and simulation frameworks (see e.g. \cite{DENAUROIS2009231}). 
It is a sufficiently adequate choice for
most applications since Cherenkov light from atmospheric particle showers
reaching the camera has a typical temporal spread of less than 10~ns, except
for the most inclined and energetic showers. The performance of the new camera readout
and data acquisition systems, however, allows full waveform sampling with an ROI
length of up to 48~samples, which is expected to increase the sensitivity of
the array to high energy showers. This mode of operation is currently being
tested on selected targets, along with more sophisticated charge integration
algorithms (see also \sref{ssec:readout_slow}).

\subsubsection{Slow Control and Connection boards}

The FPGA and ARM computer of each front-end drawer are located on the slow
control board. They are connected via a 100~Mbit/s memory bus, with a 16~bit
word width; the ARM computer has a 100~Mbit/s Ethernet interface and acts as a
device node of the distributed camera control software.
The FPGA reads out sampling data from the \nectarchip, collects other monitoring
data such as PMT currents and L0 trigger counters, and directly controls all the
electronics inside the drawer. The ARM computer runs a slow control server  
 accessing the FPGA registers, reads out all FPGA data, buffers it and sends
it over the network to a central camera server via TCP/IP using the \zmq \
library \cite{zmqsite}. The central camera server controls the drawer by means
of remote procedure calls implemented using the Apache Thrift library
\cite{apachethrift}. 
\\
The drawer slow control board also houses several point-of-load
regulators and DC line filters, providing the required voltage supplies for all
the drawer components. Also, the sockets for the PMT HV bases and the
corresponding control and readout electronics are located at the front-facing
end of the board.

The connection board has 2~RJ45 sockets, one for standard Ethernet and one for
four Low Voltage Differential Signaling (LVDS) signals: two trigger outputs, a
clock and a readout control input. A 4-pin M8~socket provides 24~V DC (see
\sref{sec:cabling}) to the main step-down (24~V to 12~V) DC-DC converter, which
is also hosted on the connection board. This arrangement assures galvanic
isolation of the electronics inside each drawer, preventing ground loops and
current surges. It also isolates the rather noisy switching-mode DC-DC
converter from the sensitive analog front-end part of the drawer, and allows it
to be efficiently cooled. The 12V output of the DC-DC converter is routed to
the regulators on the slow control board. 


\subsection{Back-end electronics}

\begin{figure}
  \centering
  \includegraphics[height=0.25\textheight]{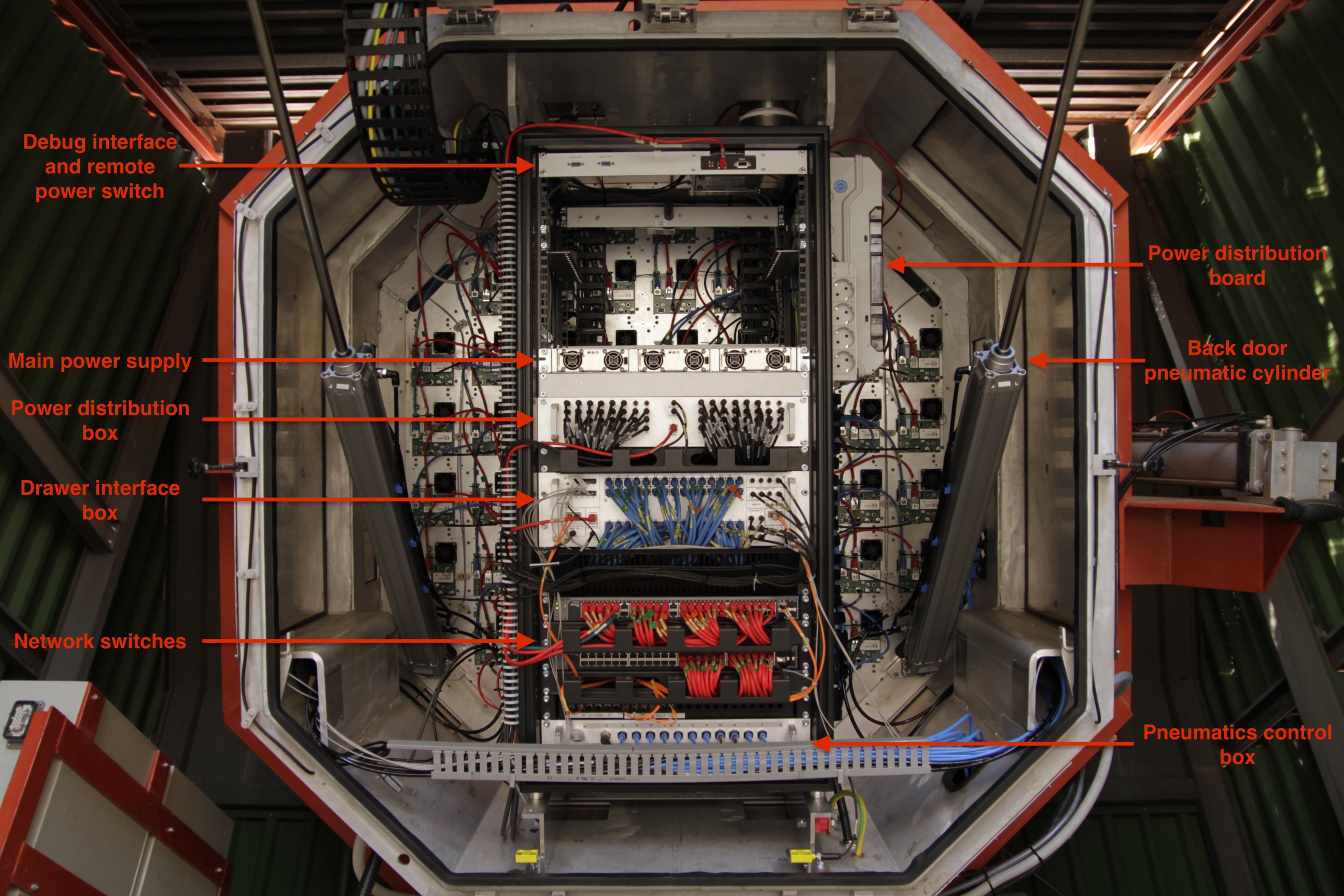}
  \includegraphics[height=0.25\textheight]{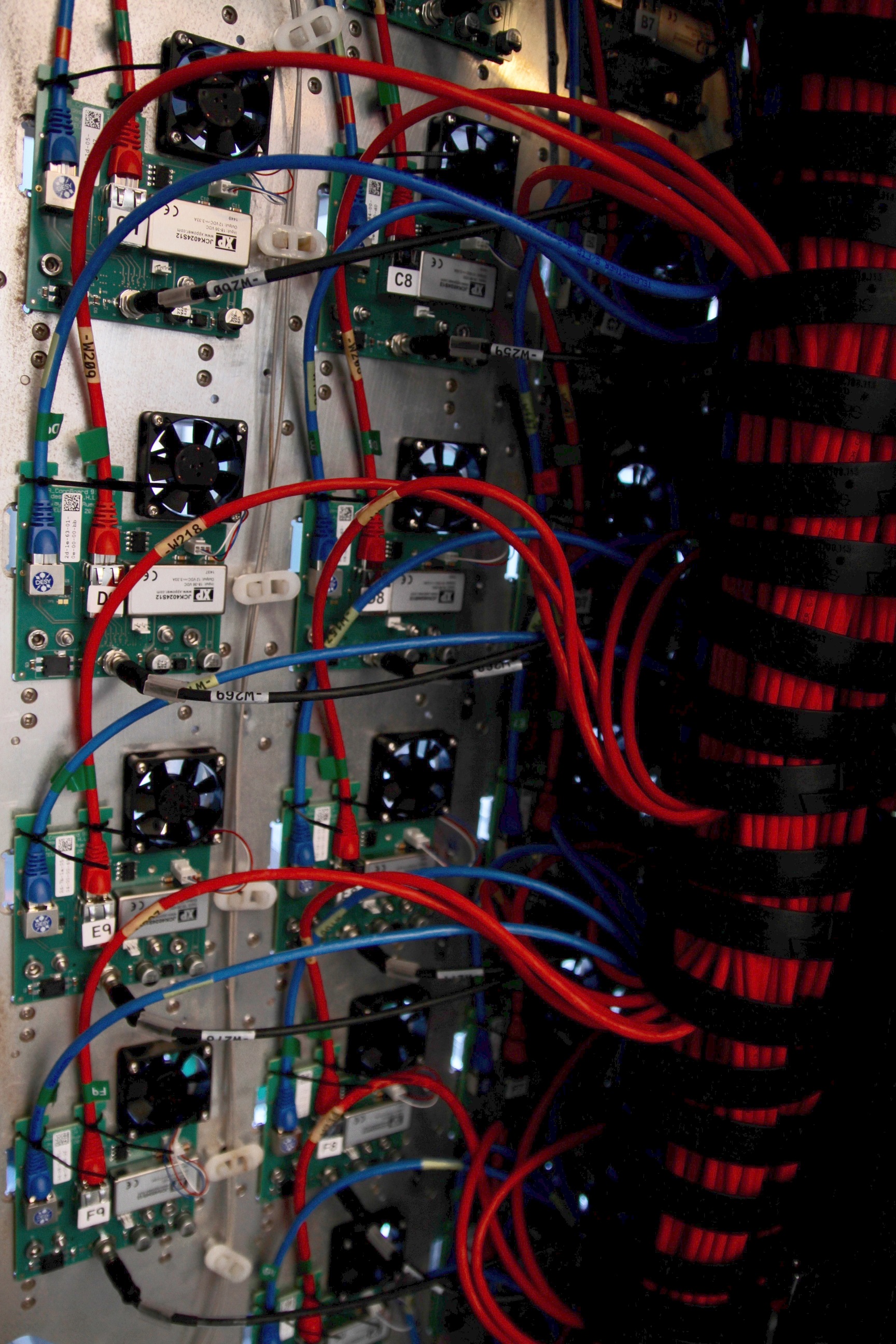}
  \caption{Left: Photograph of the inside of the installed CT1 camera. Right:
    Photograph of the camera cabling solution, using cable spines. The cables
    carry Ethernet data (red); trigger, clock and control signals (blue)
    and power (black). The cables are connected to the drawer connection
    boards.}
  \label{fig:backplane}
\end{figure}

The back-end electronics is deployed inside one 19~inch rack located in the
back side of the camera (see \fref{fig:overview}, right, and
\fref{fig:backplane}).  New components developed specifically for this upgrade
are described in the following.

\subsubsection{Drawer interface box}

The drawer interface box (DIB) is the central hub of the camera. As such, its
functions include: trigger and readout control interface and clock distribution
to the drawers; camera-level trigger generation; interface to the array central
trigger and to the auxiliary camera components, such as the front position LEDs,
the pneumatics control and the ambient light sensor (see
\fref{fig:architecture}); GPS timestamping of events and a safety interlock
logic to ensure the protection of people, PMTs and camera electronics.

\begin{figure}
  \centering
  \includegraphics[width=0.47\textwidth]{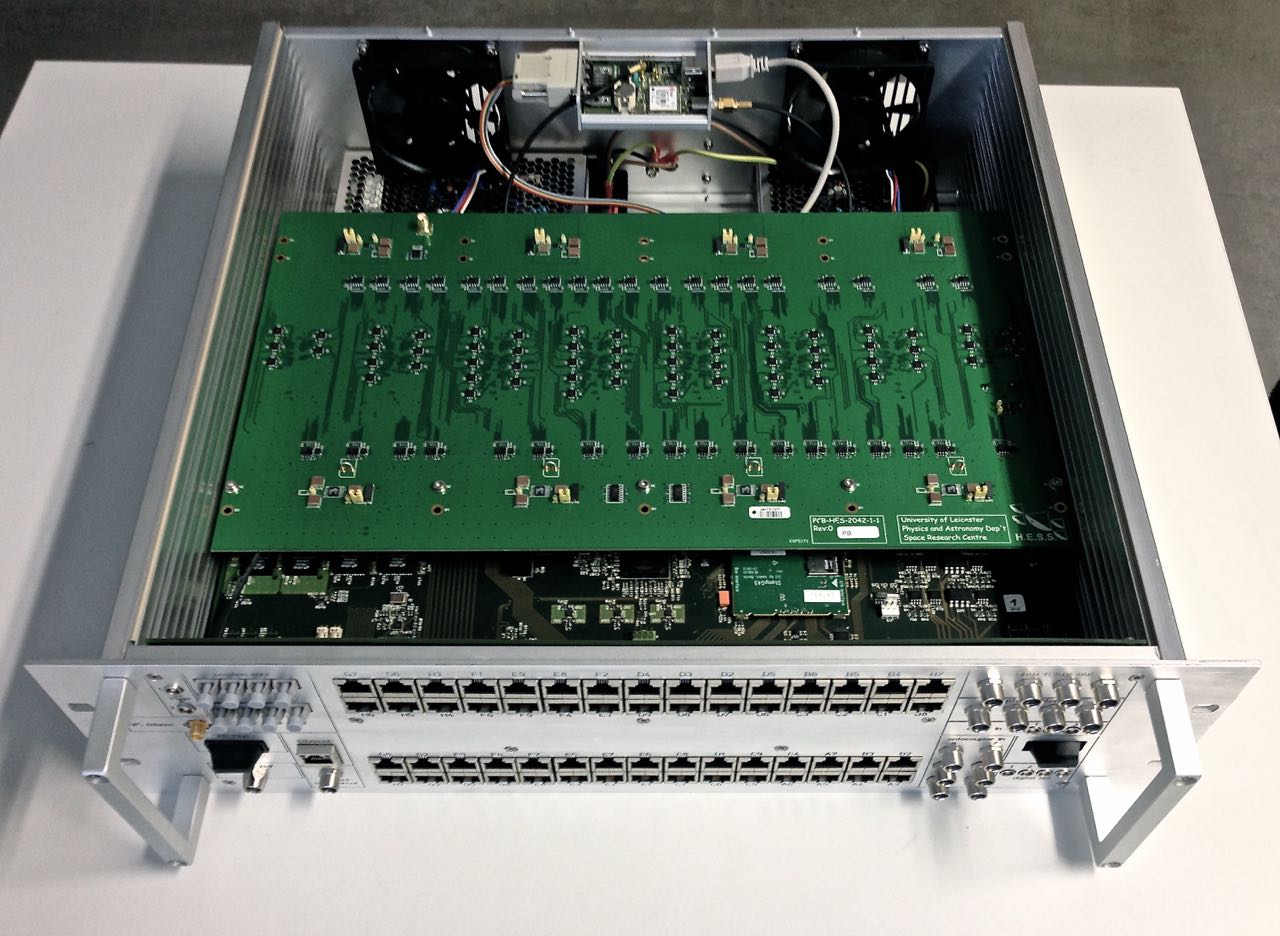}
  \includegraphics[width=0.51\textwidth]{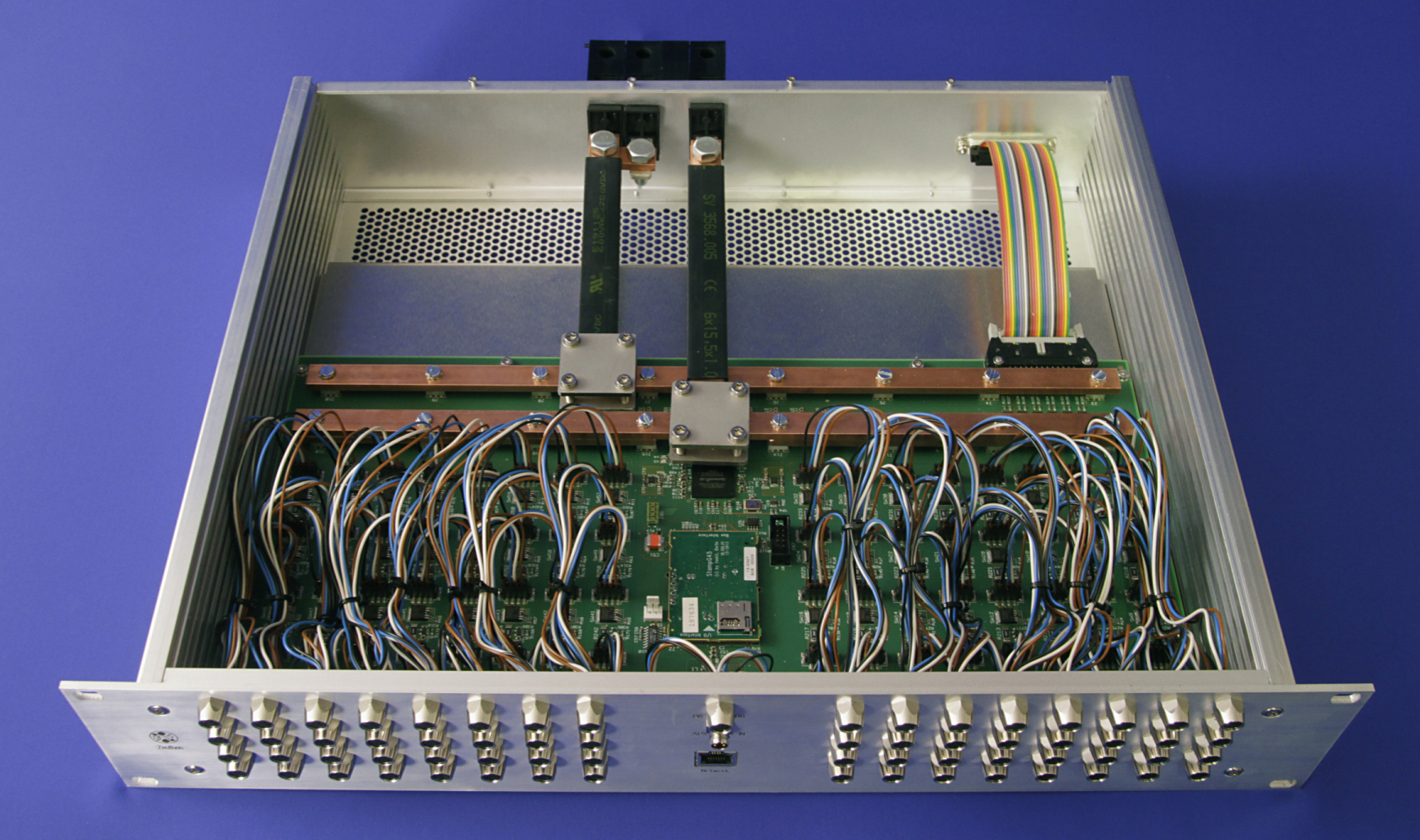}
  \caption{Left: Drawer interface box, with top lid open to show the analogue
  trigger board on top of the main board. The GPS module is at the back. Right:
  Power distribution box.} \label{fig:boxes} \end{figure}

The DIB is composed of three interconnected boards: front panel board, main
board and analogue trigger board (see \fref{fig:boxes}, left). The front panel
board houses connectors for the drawer trigger, clock and control signals, the
central trigger fiber interface, the front position LEDs lightguides and the
other camera sensors and actuators; the main board is where the FPGA and ARM
computer are located and all signals are routed; finally the analog board is a
mezzanine of the main board, whose purpose is to generate the level-1 (L1)
camera trigger (see \sref{ssec:camera_trigger}).


Furthermore, the DIB is equipped with a GPS module that delivers a
pulse-per-second (PPS) signal, to which the main 10~MHz clock, provided by a
high precision temperature stabilized quartz oscillator, is disciplined. This
clock is also distributed to the drawers. The GPS module also sends a timestamp
to the DIB via a serial interface. This is used to timestamp events at the
camera level. The precision of the camera GPS timestamp is better than few
{\textmu}s, of the order of the signal transit time between the camera and
the central trigger.

\subsubsection{Camera trigger and array trigger}
\label{ssec:camera_trigger} 
The camera trigger architecture is the same as it was in the old camera
electronics, a $N$-majority over ``trigger sectors'' of 64 contiguous pixels
\cite{trigger-rolland}. Therefore, an $N$-fold coincidence within a sector is
sufficient for the camera to trigger. Usually $N$ is set to 3.  There are
38~sectors in the camera, which overlap horizontally by one half drawer and
vertically by one full drawer.
\\
This trigger architecture is implemented as follows: the signal from of each PMT
is amplified and compared to a threshold $P$ to produce the L0 signal, which is
then routed to the FPGA on the slow control board and sampled there at 800~MHz.
The sampled L0 signal can thus be delayed or stretched in steps of 1.25~ns. Then,
the FPGA counts the number of pixels with an active L0 in each half of the
drawer separately. These two numbers are continuously sent as two LVDS
pulse-amplitude modulated trigger signals to the DIB. The amplitude modulation
has 8 discrete levels with an amplitude of 33~mV each.
\\ 
In the DIB these amplitude-modulated signals are made single-ended and
isochronally routed to 38 analogue summators, one per sector, located on the
analogue trigger board. Due to the overlapping geometry, each signal is
distributed to up to 4 sector summators. The amplitude of the output of each
summator is proportional to the number of active L0 signals in each sector. This
sector sum signal is then routed to a comparator, where a sector threshold $Q$
corresponding to $N$~active pixels is applied. All comparator outputs are
subsequently routed to the FPGA of the DIB where they are combined in an OR to
form the camera L1 trigger. When an L1 trigger is present, a length-encoded
``stop'' signal is broadcast to all drawers via the LVDS readout control lines,
and an ``active'' signal is sent to the central array trigger in the control
building via an optical fibre. Upon receiving the ``stop'' signal, the drawer
FPGA stops the \nectar\ writing, and immediately performs the readout and
digitization of the region of interest, storing the data in a front-end buffer.

During regular observations, the \hess\ central trigger \cite{hesstrigger} sends
back an ``accept'' signal to the CT1--4 cameras only if a coincidence of at
least two telescopes ``active'' signals occurs within an 80~ns window (after
correcting for their pointing-dependent light propagation delay). This signal is
received by the DIB and forwarded to the drawers, initiating there the storage
of the data held in the front-end buffer. Should no ``accept'' signal arrive,
the front-end buffer is discarded after a hold-off time $t_{b}$ slightly longer
than the readout dead-time and the maximum latency of the signal response from
the central trigger. If another L1 trigger is issued by the camera before the
hold-off is expired, a ``busy'' signal is sent to the central trigger instead,
but no signal is sent to the drawers. ``Active'', ``accept'', and  ``busy''
triggers share the same fibre connection, so they are pulse-length coded.


A design choice different from the original \hess\ camera trigger, and inspired
by the digital camera trigger design for CTA \cite{wischnewski_performance_2011}
is the 800~MHz sampling of the pixel trigger comparator output (the old logic
was asynchronous). One advantage of using a synchronous logic is that the L0
signal can be delayed and stretched, another is the possibility to implement
alternative L1 trigger logic architectures. Indeed, two of them have been
implemented: a compact next-neighbour (NN, \cite{BULIAN1998223}) logic, and a
pseudo-analogue sum trigger logic \cite{rissi_new_2009}. In both cases no
changes in the analogue part of the trigger are made, only the FPGA firmware is
different.

In the NN logic, the L1 signal is issued only when a cluster of neighbouring
pixels inside a drawer is simultaneously active. In the FPGA this is implemented
with a simple look-up table. The implementation however does not take into
account NN groups overlapping two drawers.

The pseudo-sum trigger algorithm works by measuring the duration of the L0
signals, instead of just counting the active ones. The idea behind this is that
the duration of the L0 signal is proportional to the total charge deposited
within the corresponding pixel, because for PMT-like pulses the duration above a
certain threshold is proportional to their amplitude (the pulses are roughly
triangular, see \fref{fig:gain_calib}, left). This measurement is performed in
the FPGA for each half-drawer separately, in units of 1.25~ns, within a 5-ns
window. The windowing limits the maximum contribution of any L0 signal to 4
counts, and is meant to avoid problems due to PMT after-pulsing similarly to an
amplitude clipping. The sum of the duration of the L0 signals of a half-drawer
in the preceding 5~ns is transmitted to the analogue trigger board, so the
output of any sector summator is proportional to the total charge deposited
within the corresponding sector, with an individual pixel clipping given by the
windowing.

\subsubsection{Ventilation, pneumatic and power systems}

The ventilation system consists of a single 250~mm Helios KVW~250/4/50/30 
centrifugal fan, two filters in series (coarse and fine) and a 6~kW air heater.
The whole system is attached to the back door. When operating, it forces a
$\sim360$~l/s airflow from the back to the front of the camera, where the
outlets are located. The filters ensure that very little dust enters the camera.
The heater is turned on automatically if the external humidity is higher than
75\%, to prevent condensation, or the external temperature is below
5~$^{\circ}$C, to minimize temperature gradients across the camera. In
operation, the drawer temperature is kept stable at $\sim32$~$^{\circ}$C, with a
gradient of $\pm5$~$^{\circ}$C along the top-bottom direction. Both absolute
temperature and temperature gradient have no measurable effect on the data and
on the trigger efficiency. The internal temperature of the camera is stable
for the typical range of external night temperatures, between 0 and
25~$^{\circ}$C.

The pneumatic system consists of two cylinders for the back door, one cylinder
and five clamps for the front lid. Compressed air is provided by an industrial
compressor located in the camera shelter. A custom-built pneumatics control box
implements a simple control logic using air valves. There are two modes of
operation: local or remote. In local mode, all remote operations are inhibited
and the front lid and back door can be opened manually using switches on the
outside of the camera body. In the default remote mode, only the front lid can
be opened and closed using a relay controlled by the DIB. In this mode, a power
failure or safety alarm causes the the front lid to close automatically. The
status of the pneumatic system is monitored by four sensors: a contact sensor
for the back door, two end switches for the front lid, and the remote/local
switch. An air horn is blown for a few seconds as a warning before any movement
happens.

The camera power is supplied via standard industrial 400~V three-phase AC
mains. Care was taken to ensure that the load was balanced over all three
phases. The ventilation system is directly powered by the mains, while a
distribution board provides 230~V single-phase AC to the network switches, the
front-end power supply, and the DIB. The DIB AC power is remotely controlled by
a commercial network power switch.
\\ 
The 24~V~DC to the drawers is generated by the main front-end power supply, a
commercial TDK-Lambda FPS-S1U unit, equipped with 3 load-sharing FPS1000-24
modules. It is distributed to the drawers by a custom-built power switch called
Power Distribution Box (PDB), see \fref{fig:boxes}, right. The PDB monitors the
current drawn by each drawer, samples the current and voltage ramps at power-up,
and can shut the drawers off autonomously if it detects an over-current. This
device also employs the FPGA + ARM computer design found elsewhere in the
camera. The power consumption of the whole camera is between 3 and 9~kW,
depending if the air heater is used or not.

\subsubsection{Cabling}\label{sec:cabling}
%
%
The cabling uses industry solutions such as standard Ethernet twisted-pair
cables wherever possible to ensure ease of procurement and replacement. The
data (both readout and slow control) between drawers and backplane is
transmitted  via TCP/IP over Ethernet by means of standard Cat.~6 cables. The
LVDS pulse-amplitude modulated trigger, readout control and 10~MHz~clock
signals are transmitted on standard Cat.~6$_A$ Ethernet cables of equal
length (with a tolerance of $\pm40$~mm, corresponding to $\pm0.2$~ns). The power
is delivered to to each drawer on 4-wire cables terminated with threaded 4-pin
M8 connectors. Ethernet and power cables are bundled in special pre-built cable
spines (see \fref{fig:backplane}, right). Auxiliary sensors
and actuators are connected using 3-wire or 4-wire electrical cables and standard threaded M8
connectors, except for the GPS antenna which is connected via a
standard coaxial cable and SMA connector.


\subsection{Auxiliary and calibration devices}
Several sensors are deployed inside and outside the camera to monitor door
position, temperature, humidity, ambient light and smoke presence. Their
signals are fed to a safety interlock system that ensures safe camera
operations for both shift crew and hardware. The interlock logic is implemented
in the firmware of the DIB FPGA, so it cannot be disabled and is independent of
the software implementation.

For the calibration of the gain of the individual PMTs, a device called single
photo-electron (SPE) unit is used. It is located in the shelter, facing the
front of the camera. Due to its position, it can be used only when the
telescope is fully parked. The SPE unit uses an LED to emit pulses
of blue (370~nm) light with pulse frequencies ranging from 38~Hz to 156~kHz.
The intensity of the pulses ranges from $\sim0.1$ to $\sim200$~photo-electrons,
and their duration is less than a nanosecond. It was designed at the LPNHE
laboratory in Paris for the original \hess\ array. A plastic diffuser in front
on the LED ensures complete camera illumination, with a uniformity of about 50\% \cite{aharonian_calibration_2004}.
The pulse frequency and intensity are controlled by the camera server via UDP.
An adapter board was added to the SPE unit, allowing it to send its trigger
signal to the camera via an optical fibre connection. This additional trigger
signal is synchronous to the light pulses and it is required for calibration
purposes (see \sref{ssec:gainff}).
\\ 
To perform the pixel-wise calibration of the light collection efficiencies of PMT photo-cathode
and funnels \cite{BERNLOHR2003111}, another device called flat-fielding unit is used
\cite{aye_novel_2003,aharonian_calibration_2004}. It is located in the centre of
the telescope mirror dish and, similarly to the SPE unit, it has a LED that
emits short ($<3$~ns FWHM), blue (390--420~nm) light pulses of fixed intensity
($\sim100$~p.e. at the PMTs). A holographic diffuser is placed in front of it.
The high quality diffuser and the small angle subtended by the camera assure a
uniform illumination. The stability of the flat-fielding intensity and its
non-uniformity across the camera are within 5\% RMS.

\subsection{Camera network and software}
%

All main subsystems (drawers, DIB, PDB and ventilation system) are connected via
100~Mbit/s Ethernet links to two interconnected 48-port switches located inside
the camera. Their uplink to the main camera server is a 10~Gbit/s optical fibre
connection. The camera server is a commercial 1-unit rack server with a 4-core
Intel Xeon E3-1246v3 processor clocked at 3.5~GHz and 16~GB of DDR3 RAM. It is
housed in the computer ``farm'' (see \fref{fig:architecture}), an
air-conditioned server room inside the main control building. The topology of
the internal camera network is star-like: slow-control commands are issued only
by the central camera server, which is also the endpoint of the monitoring,
logging and event data streams. The devices on the network are independent from
one another, and the only access point to the camera is through the camera
server. \\
This distributed design improves the flexibility and resilence of the camera:
for instance, during data-taking the ARM computer memory (256~MB per module, for
a total of 15.6~GB) is used for buffering the data, preventing data loss during
event bursts.

The software was written from scratch; it has a
distributed, multi-architecture nature, as required by the new camera design.
Its main functions are slow control and event acquisition; it also includes
text-based and web-based user interfaces, extensive unit tests, integration
tests and validation routines needed for the mass production, and a full
commissioning and calibration suite able to take runs, analyze them and adjust
camera parameters independently of the main \hess\ DAQ.

The full codebase is around 100,000 lines of code long, composed by 82\% C++,
11\% ANSI-C and 7\% python. Its implementation was one of the major efforts of
the upgrade, and required around 6 man-years by a team composed of two full-time
coders and four part-time contributors. This paid off with a 10- to 1,000-fold
improvement in speed and reliability over the previous system (see
\sref{ssec:readout_slow} for some performance measurements). 

To maximize efficiency, extensibility and maintainability of the codebase, the
development team made use of well-tested off-the-shelf open source solutions
wherever possible. A single source tree was used for both ARM and x86\_64
architectures; cross-compilation was handled by the CMake build system. The
operating system running on the ARM computers is the Yocto embedded Linux
\cite{yoctoproject}. It runs a Linux kernel v3.0 patched by the manufacturer,
and a custom-built DMA-enabled driver for communicating with the FPGA. The
remote procedure call framework required to control the camera is implemented
using the Apache Thrift library. The camera slow control software was interfaced
to the already existing \hess\ data acquisition software (DAQ, \cite{hessdaq})
via the CORBA protocol. Data transfer is accomplished via the \zmq \
\cite{zmqsite} smart socket library. The raw data serialization protocol is
custom, and optimized for speed; for monitoring and logging the general-purpose
Google Protocol Buffers library \cite{protobuf} is used instead.


\section{Test facilities and procedures}
\label{sec:test}

The development of a new detector generally requires planning and implementing
test and verification procedures. The testing needs of \hessIu\ cameras were
identified early on in the project and grouped into four main stages
(prototyping, integration, quality control, commissioning), for which four
distinct test facilities were build, and are described in the following.



\begin{figure}
  \centering
  \includegraphics[width=0.95\textwidth]{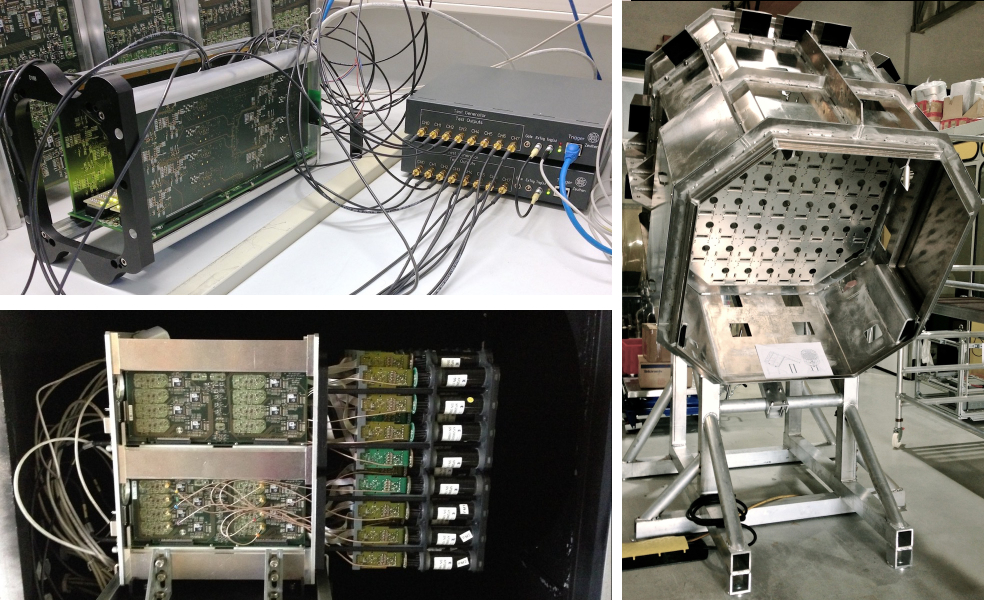}
  \caption{Left, top: table-top drawer test-bench used for the quality control
  of the mass-produced drawers. Note the daisy-chained pulse generators stacked
  one on top of the other. Left, bottom: mini-camera. Right: full copy of the
  camera body used for testing at DESY in Zeuthen. The camera inclination
  reproduces the parking position in Namibia: this helped training the
  deployment. The mechanical structure was fabricated at the LLR laboratory in
  France.}
    \label{fig:copycam}
\end{figure}

\subsection{Table-top laboratory test bench} 

During the prototyping stage daily debugging and testing of the prototypes was
needed to validate the design of the new electronics and the correct
implementation of all features needed. These mostly manual and one-time
characterization tests required a versatile laboratory test setup.
\\
For this purpose, a table-top laboratory test bench was set up, equipped with
an oscilloscope (LeCroy~DPO~4104), an arbitrary function generator
(Agilent~81160A), and several auxiliary instrumentation such as a variable
attenuator and a digital multimeter. Many results shown here, such as the
linearity shown in \fref{fig:linearity}, left or the bandwidth shown in
\fref{fig:deadtime}, right, have been obtained using this setup.

When the mass production of 270 drawers started, each one of them had to
undergo more than 300 individual tests to pass the quality control. The tests
mainly checked the functionality of the drawer, but also included the
calibration of the \nectarchip; and the characterization of readout noise,
linearity, saturation, cross-talk.
\\
The table-top test bench was thus refitted with four purpose-built,
Ethernet-controlled 8-channel pulse generators, allowing to perfom the
above-mentioned tests automatically. The generators are built using the same
FPGA-ARM computer combination used elsewhere and were seamlessly integrated in
the test software. They deliver PMT-like pulses with fast ($\sim 1$~ns) rising
and falling edges, variable amplitude (from 0.6 - 300~mV), delay (0 - 64~ns in
steps of 0.25~ns) and width (2-62~ns in steps of 0.25~ns). They have a RJ45
socket to provide the LVDS drawer clock and acquisition control signal, and to
test the outgoing drawer trigger signals; an external trigger input, output and
gate, which allow them to be daisy-chained, see \fref{fig:copycam}, top left.
Two daisy-chained generators can simultaneously test all 16 channels of one
drawer: during the quality control, the test bench could therefore test two
drawers in parallel.
\\
Using this setup, running all the tests needed for quality control of a drawer
took less than 30~minutes. Several of these tests only use self-contained
testing functionalities of the drawer, such as the possibility of inject fake
pulses at the PMT signal inputs to test the trigger path. Such tests can
therefore be run even when the drawer is not on the test bench: during the
commissioning of the cameras, they proved to be an invaluable troubleshooting
tool.

\subsection{Single drawer black box test bench} 

The integration of the new front-end electronics with the existing \hessI\ PMTs
was a critical step, and it required a setup to test a PMT-equipped drawer with
a low level of background illumination, with the possibility of flashing it with
Cherenkov-like light pulses.
\\
A single-drawer ``black box'' test bench was built for this purpose. It
consists of a simple aluminium box holding a complete drawer. A \hessI\ SPE
unit is used to illuminate the PMTs and is attached to the side of the box
facing them. The inside of the box is painted black to minimize reflections.
The black box was used extensively during the first stages of prototyping, and
later on to devise the appropriate calibration routines. After prototyping was
over, it was shipped to the \hess\ site, where it was used during the
deployment of the cameras, mostly to inspect malfunctioning drawers during the
day. It is still being used on site occasionally for drawer maintenance and
refitting.

\subsection{Mini-camera}

The verification tests needed during the integration and commissioning phases
called for a fully functional camera. A 4-drawer ``mini-camera'' was built for
this purpose (\fref{fig:copycam}, bottom left), housed in a 1~m$^{3}$
light-tight enclosure, with 64~PMTs, one DIB and a light source (an SPE
unit). With it, it was possible to test the integration between front-end and
back-end by recreating a minimal 1-sector trigger setup. This allowed testing
the analogue trigger board and the other trigger functionalities of the DIB
using realistic signals, as close to the field conditions as possible. The
mini-camera was also used to develop and test the slow control and event builder
software and to test their integration into the existing \hess\ DAQ control
software.

After the installation of the first telescope, the mini-camera  served as the
main commissioning test bench. In fact, it became the primary way of reproducing
and troubleshooting in the laboratory problems found during the first months of
field operations.

\subsection{Full camera copy}

Later in the project, the camera on-site assembly and integration had to be
prepared and rehearsed as thoroughly as possible before actual deployment. This
stage required a full camera, so a copy of the camera body was fabricated at the
LLR laboratory and installed at DESY Zeuthen (\fref{fig:copycam}, right). Due to
its size, it could not be housed in a light-tight room, so the PMTs were not
used, but all other components of all four cameras were mounted and tested first
on this camera body, with the purpose of verifying their functionality and
training the technicians involved in the assembly.  Thanks to this, the on-site
physical assembly of one camera could be finished in less than 5 working days.
In 2015, the total down-time of the CT1 telescope, excluding commissioning and
fine-tuning, was 18 days. In 2016, the CT2-4 telescopes had a total down-time of
four weeks.

Other testing and validation activities performed on the copy camera included
checks of the cable mapping; full trigger chain functionality check; assessment
of the event builder performance and its integration with the \hess\ software;
configuration of the camera-internal network; evaluation of the capabilities of
ventilation system, slow control software, and power supply; mechanical integration
of the new back door and the pneumatic system.


\section{Camera calibration}
\label{sec:cal}

The following section is an overview of the calibration procedures needed to
commission the upgraded cameras, partly updating the information found in
\cite{aharonian_calibration_2004}.

\subsection{Readout}
\subsubsection{Nectar line correction}
\begin{figure}
  \centering
  \includegraphics[scale=0.29]{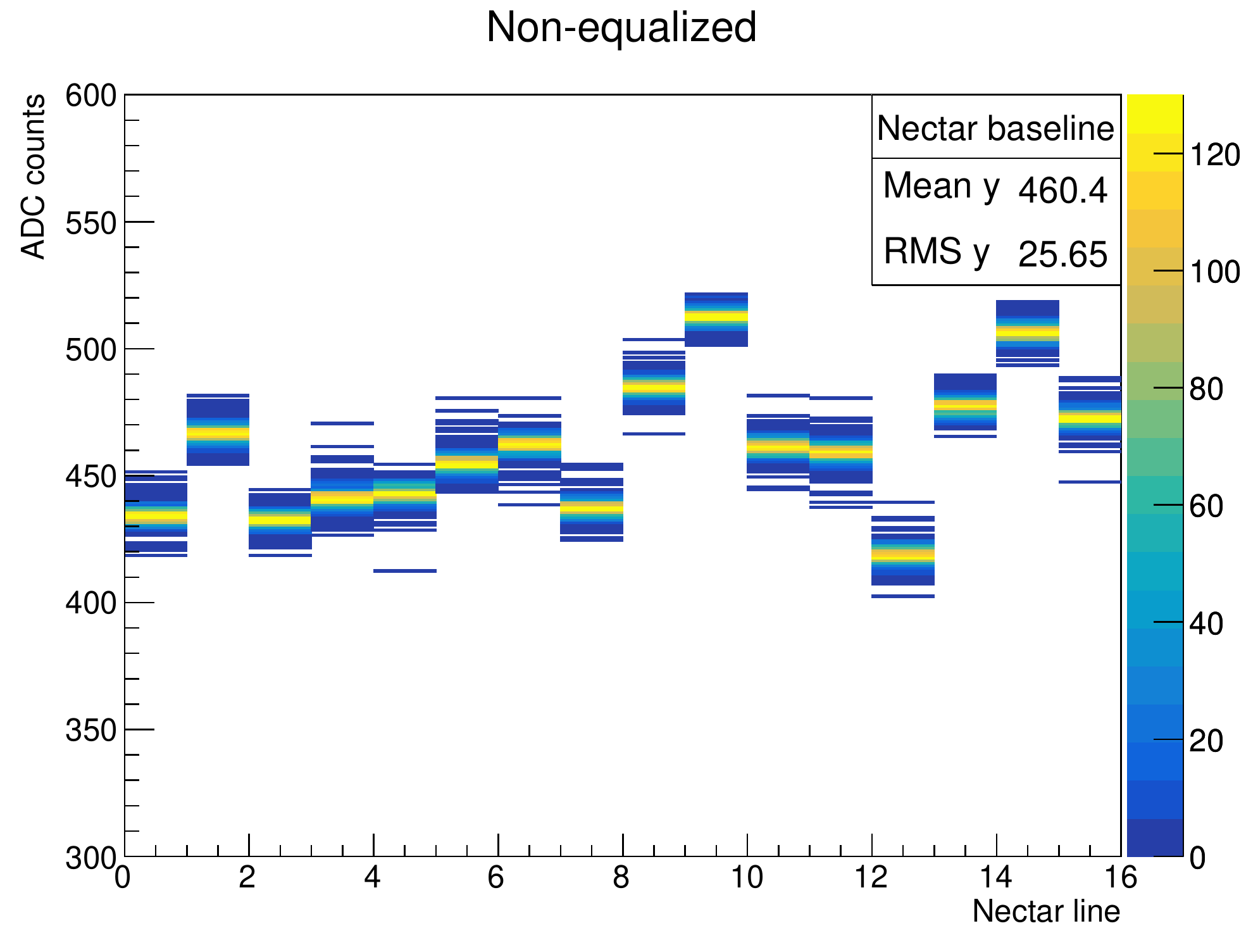}
  \includegraphics[scale=0.29]{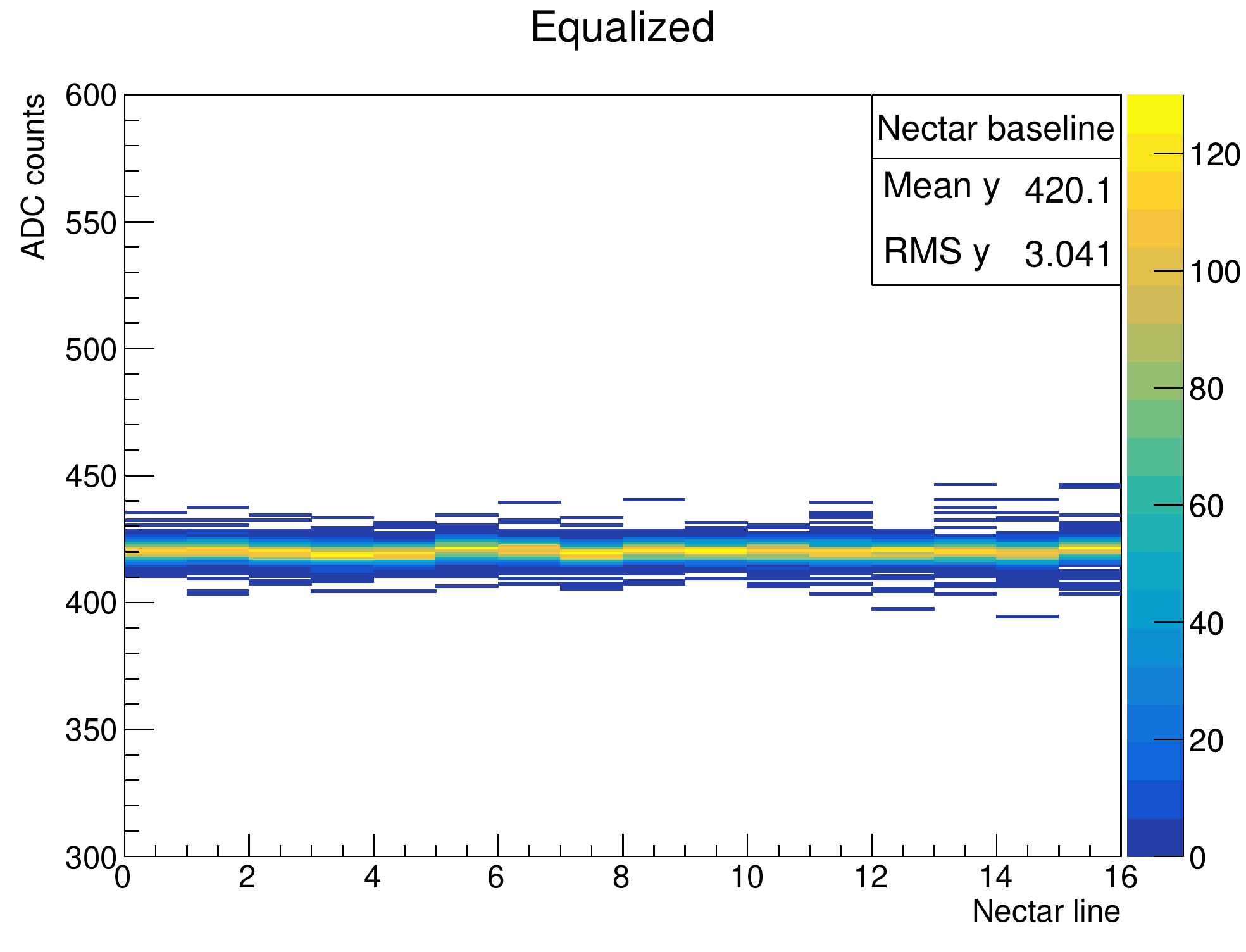} 
  \caption{Equalization of the \nectar\ switched capacitor array baseline to a
    nominal value of 420 ADC counts. The plots show the histogram of \nectar \ 
    cell readout values ordered by line before (left) and after (right)
    calibration.  Both offset and RMS are adjusted using the line DACs. The 
    final RMS of $\sim3$~ADC counts is lower the average RMS of $\sim4$ ADC counts because this data comes from a low-noise back channel on the drawer (see \sref{sec:perf_analogue}). }
    \label{fig:baseline} 
\end{figure}

The \nectar\ switched capacitor array is arranged in 16 lines $\times$ 64
columns of analogue storage cells. The analogue input is buffered by amplifiers
providing the signal for each of the 16 lines. 
Since each line has its own input buffer and readout amplifier, a spread of the
DC baseline levels between lines is observed. This dispersion is stable in time
and can be compensated by DC line offsets regulated by 16 integrated digital to
analogue converters (DACs).
In order to calibrate their values, sample-wise pedestal runs are taken with a
readout width of at least 16 samples, and the average difference between the
sample baselines and a reference default value (defined as 420~ADC counts to
reserve roughly 10\% of the dynamic range for pulse undershoots) are calculated.
The line offset DAC settings can thus be corrected to compensate this
difference. By repeating this process several times, the default baseline offset
is approached.  In only 5 iterations, the RMS of the baseline decreases from $\sim20$
to $\sim4$~ADC counts on average, see \fref{fig:baseline}. This procedure needs
to be performed in principle only once, but in practice it is performed for the
whole camera each time a drawer is exchanged, or the PMT gain is re-adjusted, so 
about twice per year on average.
The values of the line DACs are then permanently stored in a MySQL database.
This correction is not strictly necessary for regular observations, where the
only measured quantity is the charge integrated over 16 samples, minus a
time-averaged pedestal baseline. This is because the width of the ROI is a
multiple of the line dispersion period, so any event-wise baseline shift due to
it is cancelled out.

\subsubsection{Readout window adjustment} 
The \nectarchip s continuously store signals inside their analogue memory ring
buffers until the arrival of an L1 trigger signal. When this happens, the
region of interest is located $L$ cells before the last sampled one, where $L$
is the L1 trigger latency in nanoseconds. It is therefore necessary to measure
the trigger latency $L$ for each chip and trigger source. This is done by
illuminating the whole camera with high intensity ($\sim100$~p.e.) reference
light pulse (see e.g.  \fref{fig:gain_calib}, left) while varying the \nectar\
register $Nd$ controlling the start of the region of interest inside the chip
buffer, until the sampled pulse signal is located at the center of it. Since
the chip buffer is 1024 cells deep and circular, $Nd$ is the complementary of
$L$ over the buffer length, $Nd = 1024 - L$.
\\
The position of the readout window needs to be adjusted individually for two
trigger sources having different latencies: the SPE unit trigger and the
standard camera level 1 trigger.  For the former, the SPE unit itself provides
the reference light pulses; for the latter the flat-fielding unit is used.
After a successful adjustment, the two sets of $Nd$ values are stored in a
MySQL database.

\subsection{PMT gain flat-fielding}
\label{ssec:gainff}
\begin{figure}
  \centering
  \includegraphics[height=.22\textheight]{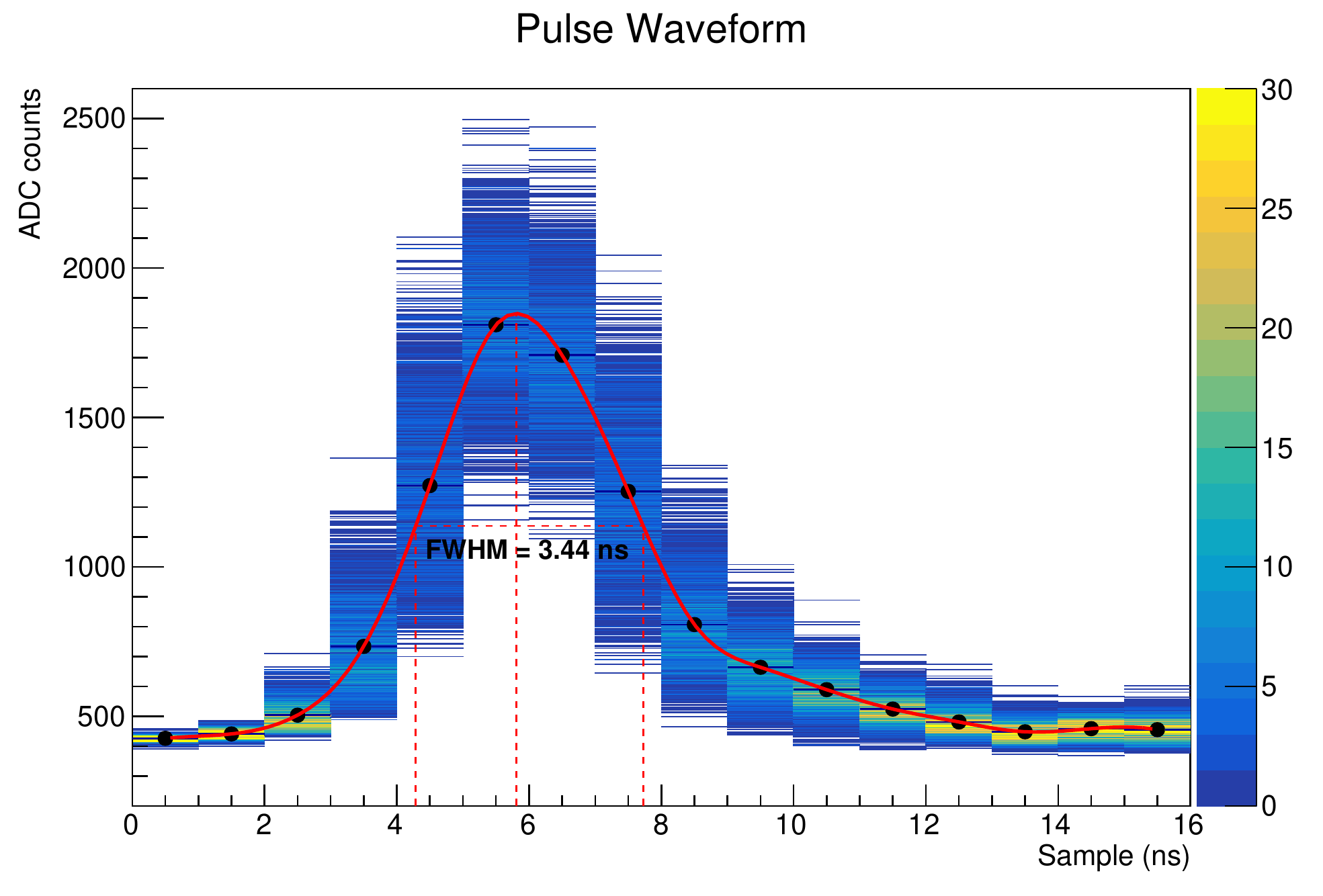}
  \includegraphics[height=.22\textheight]{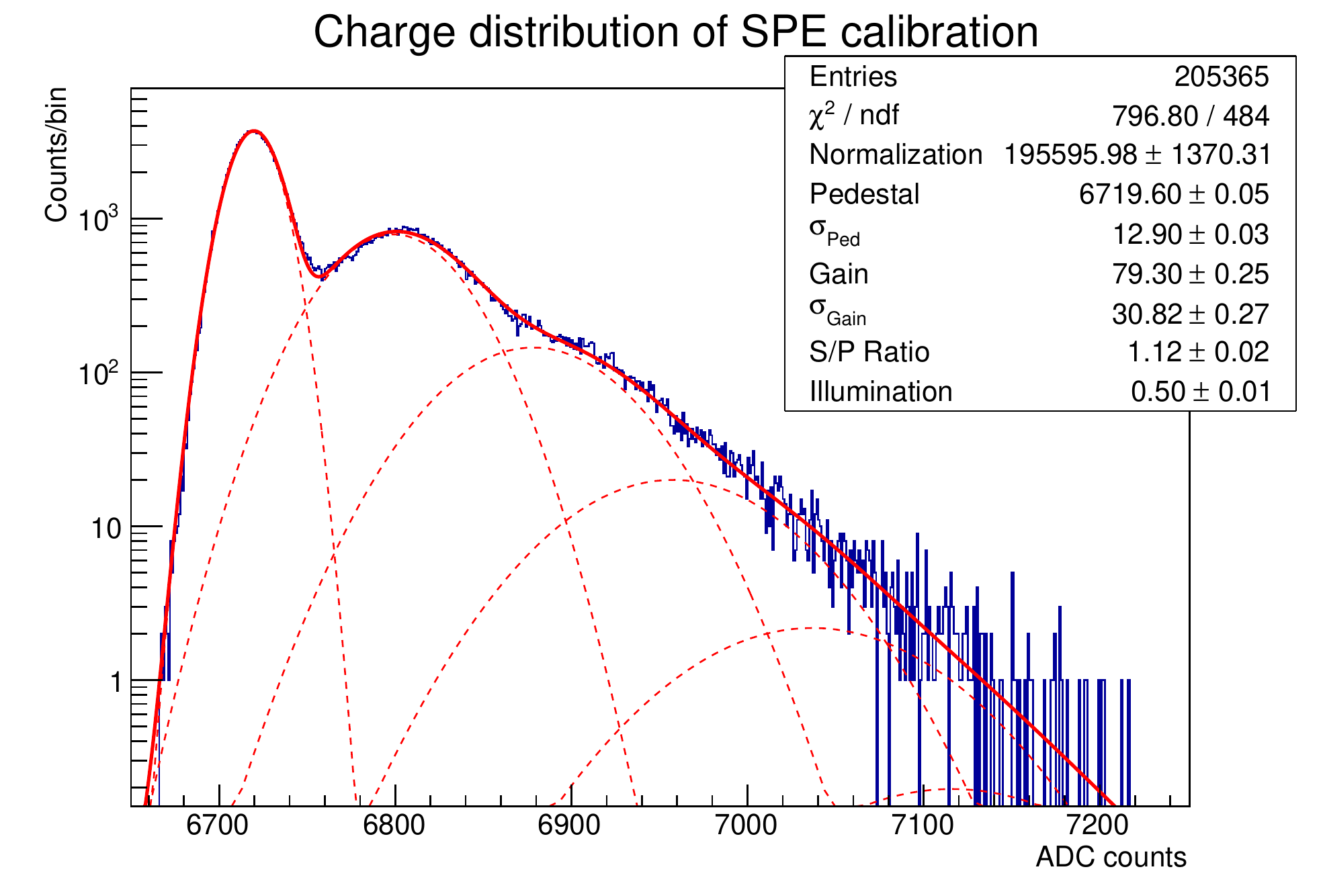}
  \caption{Left: The digitized PMT pulse from the flat-fielding unit as recorded
  by the readout. The plot shows the distribution of over 2,000 light pulses. The
  red line is a spline interpolation of the average values in each sample. The
  FWHM of this interpolation is also shown with dashed lines. Right: The
  distribution of charges from a typical PMT gain calibration run, fitted to a
  linear combination of Gaussian functions as described in
  \cite{aharonian_calibration_2004}, section 6.2, equation 6. In this particular
  case, $\gamma^{ADC}_{e}$ (``Gain'') is 79.3 and $\sigma_{P}$ is 12.9~ADC
  counts, corresponding to a gain of $2.7\times10^5$ and a noise level of
  0.16~p.e. }
    \label{fig:gain_calib}
\end{figure}

In order to reliably measure the amount of light arriving at the camera, it is
necessary to equalize the gain of the electronic chain of each channel. This is
done by varying the voltage applied to the PMTs, and illuminating the camera
with pulses from the SPE unit, at an intensity so low that the average number
of photons detected by a PMT for each light pulse is less than 1. The typical
charge distribution of these calibration runs can be seen in
\fref{fig:gain_calib}, right. The charge is integrated over the standard 16~ns
ROI. This distribution can be fit by a linear combination of Gaussian
functions, as shown in \cite{aharonian_calibration_2004}, equation~6. This
simple fit form is quite robust over a wide range of PMT illuminations (0.1--3
p.e.), but its result is biased: the actual PMT single photo-electron charge
distribution is not a Gaussian, but an asymmetric distribution skewed towards
lower values. So, the average single photo-electron amplitude is lower than the
amplitude at the peak, as shown in \cite{bernlohr_simulation_2008}. This
discrepancy is corrected later on in the analysis by a factor 0.855 derived
from realistic simulation of the \hessI\ PMTs \cite{simu_internal}. After this
correction, the systematic error in determining the PMT gain has been estimated 
with simulations to be within $\sim5$\%. Other techinques for the single
photo-electron calibration, such as those described in
\cite{Saldanha:2016mkn,takahashi_technique_2018}, are currently being evaluated.

Similarly to their predecessors, the PMT gain of H.E.S.S.~I upgrade cameras is
flat-fielded to a conversion factor $\gamma^{ADC}_{e}$ of 80~ADC counts (peak
value obtained from the above-mentioned fit). This particular value is chosen
empirically, based on the reproducibility and robustness of the fit results.
The corresponding average PMT gain is $2.72\times10^5$. The PMT voltages range
from $\sim850$~V up to $\sim$1,350~V, and are stored in a MySQL database. To
achieve a precise gain flat-fielding (better than 4\%), the procedure is
iterated several times, each with a finer voltage step. The performance of the
PMTs degrade with time, so their gains need to be flat-fielded every six
months, adding on average 13~V to the PMT voltages. After gain flat-fielding,
 the position of the region of interest $Nd$ is readjusted, because
the PMT transit time $t_{p}$ depends on the voltage $V$ applied to it.

\subsection{Camera flat-fielding}
After equalizing the PMT gains, to correctly estimate the amount of Cherenkov
light reaching the detector plane, one needs to calibrate the differences in
light collection efficiency for each pixel. As mentioned previously, this is
achieved by recording light flashes generated by the flat-fielding unit. In
fact, assuming the flat-fielding light is homogeneous, one can easily calculate
a correction factor $C_{i}$ from the charge $Q_{i}$ recorded by each pixel and
its average over all camera pixels $\bar{Q}$: $C_{i} = \bar{Q_{i}}/Q_{i}$. This
is done for the high and low gain channels separately. Flat-fielding runs are
also used to calibrate the time of maximum information of each pixel, under the
assumption that the flat-fielding pulses arrive isochronally at their entrance
window. The flat-fielding is performed several times per observation period (one
lunar month), and the obtained coefficients for the period are then averaged and
stored in a MySQL database. As for the previous cameras, the  distribution of the
$C_{i}$ coefficients is a Gaussian with an RMS of 10\%, there is no discernible
gradient across the camera. Trials with a new flat-fielding unit are ongoing.


\subsection{Trigger}

As described in section \ref{ssec:camera_trigger}, the camera trigger  has
several parameters which require dedicated calibrations. The most important ones
are the pixel and sector thresholds $P$ and $Q$, and the pixel L0 delay $d$. The
L0 stretching $l$ is set to zero to have a trigger response similar to the one
of the old cameras. Central trigger delays also have to be adjusted after the
installation of a camera.

\begin{figure}
  \centering
    \includegraphics[width=.49\textwidth]{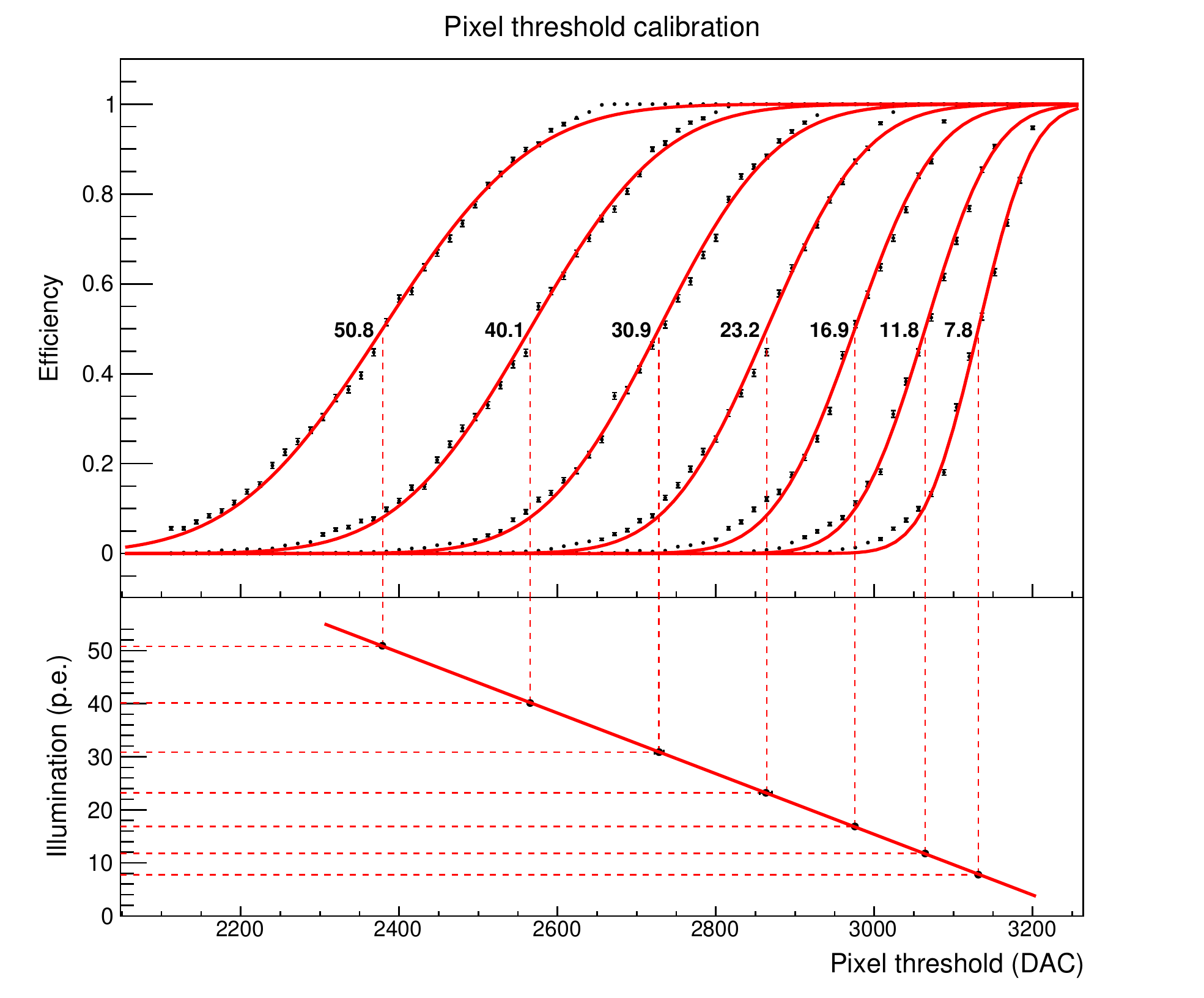}
    \includegraphics[width=.49\textwidth]{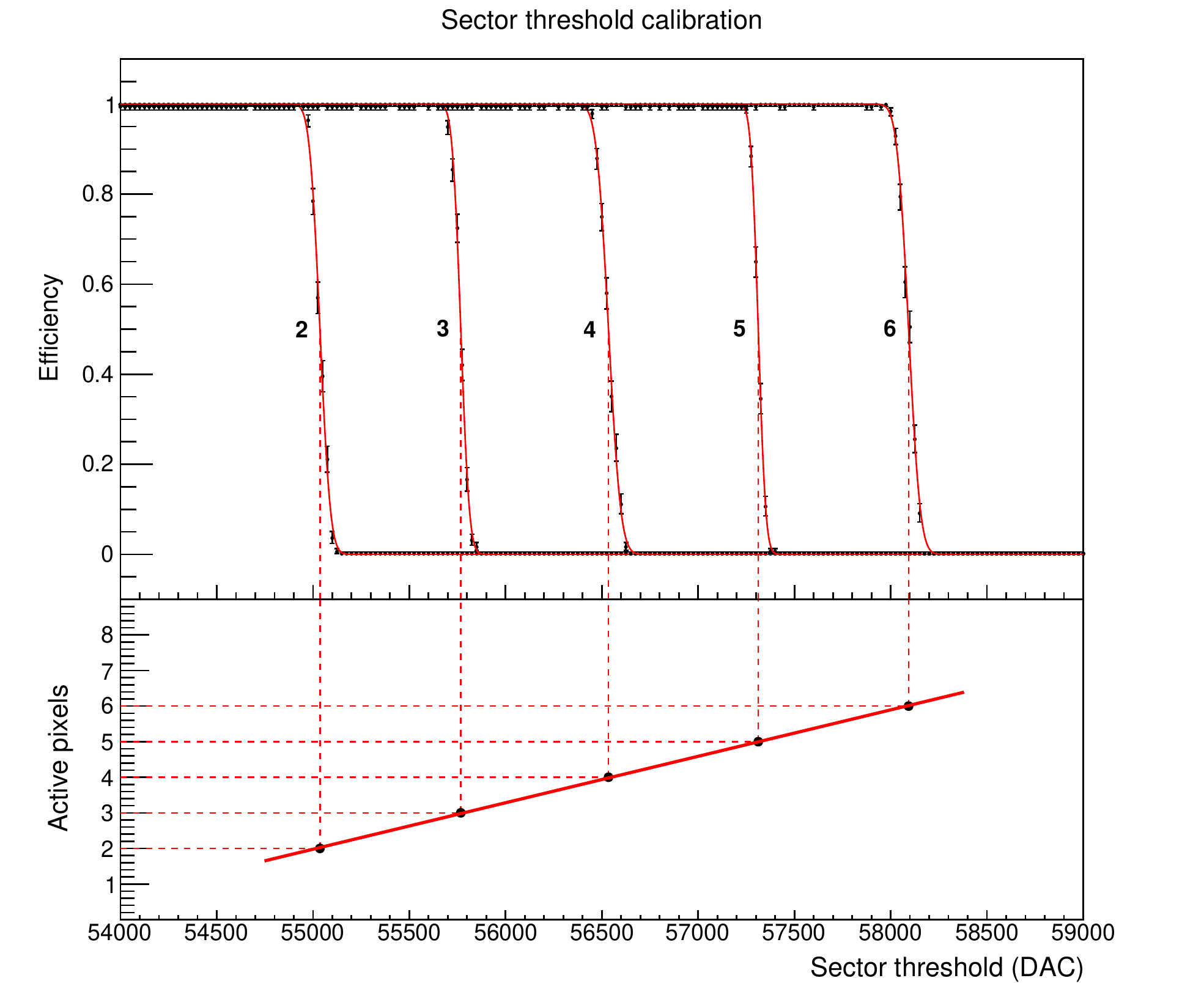}
  \caption{Efficiency curves from pixel (left) and sector (right) threshold
    calibration. The value of the pixel and sector thresholds $P$
    and $Q$ are given in units of DAC counts (0.76 mV/count). 
    Lower insets show a linear dependency of the sigmoid center on the illumination level (in p.e.) and the number of actively triggering pixels for pixel and sector thresholds, respectively. The error bars
    on the efficiencies are calculated following approximation 1 of
    \cite{casadei_estimating_2012}. }
    \label{fig:sectorthr}
\end{figure}

Calibrating the pixel threshold $P$ requires finding the relationship between its
value as set by the electronics, in DAC counts, or mV, and its effective value
in photoelectrons. This is done with special calibration runs, where a
variable-intensity pulsed light source is needed. The SPE unit is used for this
purpose. The camera is flashed with a fixed frequency $f$ and an varying
intensity $I$, between $\sim5$ and $\sim50$ p.e. The light pulse intensity in
each pixel is measured from the mean of its charge distribution using a
previously-determined PMT gain coefficient. While the run is ongoing, $P$ is
varied and the L0 pixel trigger efficiencies are measured as the ratio between
the pixel trigger rate and $f$. The resulting graph is a sigmoid, whose
mid-point marks the value of $P$ needed to discriminate $I$ photoelectrons (see
figure \ref{fig:sectorthr}, left). By repeating this procedure for several
intensities, it is possible to determine the offset $b$ and slope $m$ of the linear
dependency $P(I) = mI + b$, and thus the effective value of $P$ in
photoelectrons.

The calibration of $Q$, the sector threshold, is likewise accomplished by means
of special flat-fielding runs. With the flat-fielding unit activated at a
frequency $f$, one enables $N$ pixels in every sector, varies $Q$ and measures
the sector trigger efficiency as ratio between the measured sector trigger rate
and $f$.  The mid-point of the resulting sigmoid curve corresponds to the value
of the threshold $Q$ for the given $N$ (see figure \ref{fig:sectorthr}, right). Since
$N$ is discrete, this sigmoid is much steeper than the one for $P$. By repeating
this procedure for several values of $N$, it is possible to determine the
relationship $Q(N)$ in a similar way as for the pixel threshold. 
Finally, for the nominal multiplicity of 3, $Q$ is set to a value for which all
sectors have 100\% efficiency when $N=3$ and 0\% when $N=2$.

The L0 delays calibration is much simpler: using a modified drawer FPGA
firmware, it is possible to send the sampled L0 information of all pixels on
the data stream. This is done while flashing the camera with the flat-fielding
unit, so all pixels are illuminated at the same time. The L0 delays $d$ are
then individually adjusted until the rising edge of all L0 signals is aligned.

\begin{figure}
  \centering
  \includegraphics[width=\textwidth]{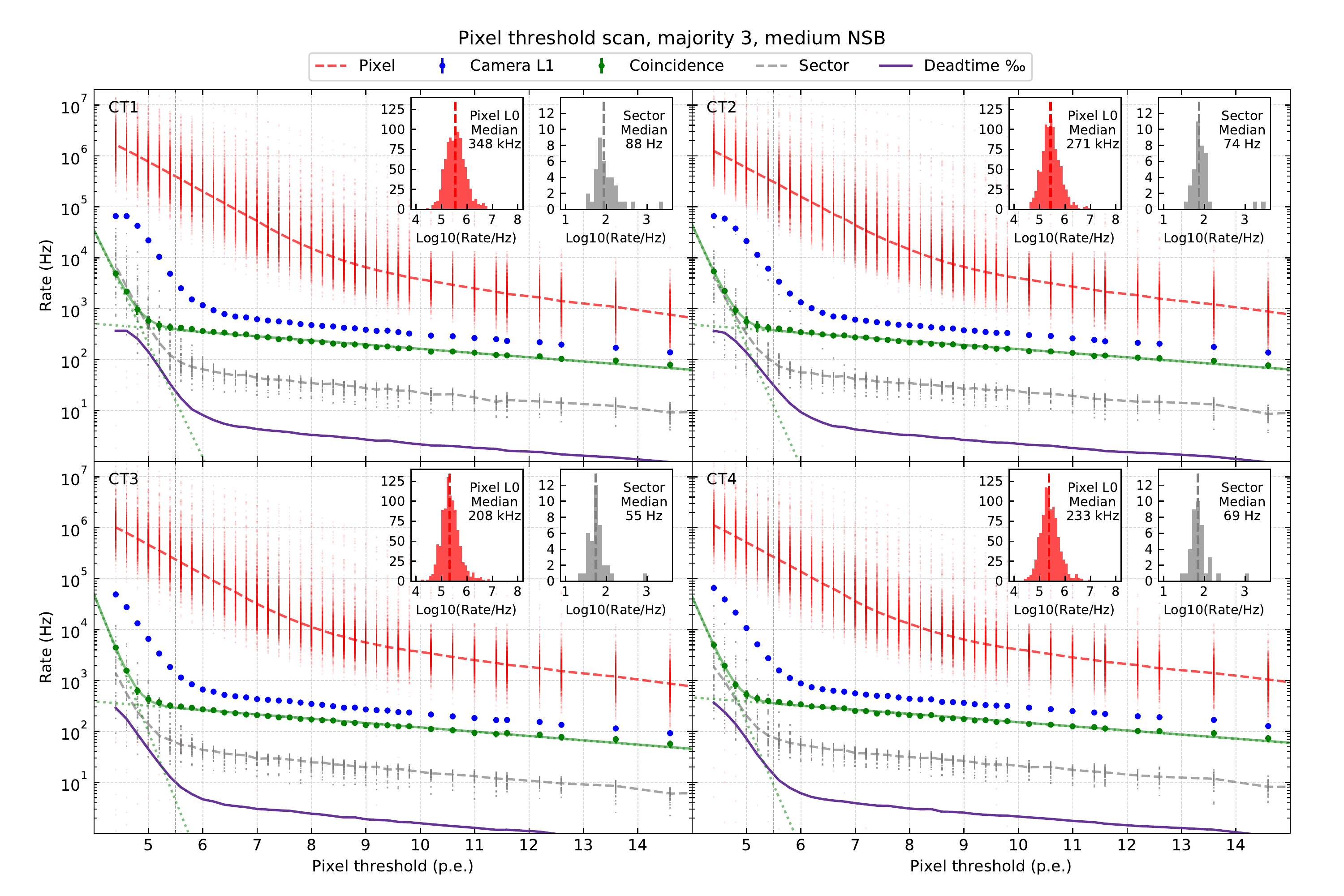}
  \caption{Results of a threshold scan for a Galactic source with a typical
    level of NSB light. The graphs show trigger rate
    versus pixel threshold in photo-electrons. Pixel and sector rates are
    showed alongside camera L1 trigger rates and coincidence trigger rates with
    any other telescope in the array, including CT5. The coincidence trigger is
    formed after applying delays dependent on pointing direction. The green
    line is the result of the fit of a linear combination of two exponential
    functions to the coincidence rate data, the red and grey dashed lines show
    the median rates of pixels and sectors, respectively. The fraction of
    events lost due to dead-time is shown in per~mille as a purple line. The cameras
    are operated at a nominal threshold of 5.5 p.e, shown as a vertical grey
    dashed line. Histograms of the rates of all pixels (red) and sectors (grey)
    at this nominal threshold are shown in the insets }
  \label{fig:ff_pixscan}
\end{figure}

After the above-mentioned calibrations, it is necessary to determine the
operating point of $P$. To do that, $P$ is varied while measuring the camera L1
and coincidence trigger rates during a regular observation run. This ``threshold
scan'' is performed under optimal observing conditions, using the whole array,
including CT5. It results in ``bias curves'' for all four telescopes, shown in
\fref{fig:ff_pixscan}. In these plots, the steeply falling part of the
coincidence rate at thresholds lower than $\sim5$~p.e. is due to noise from the
night sky background (NSB) light, whereas the flatter part at higher threshold
values is due to cosmic-ray showers. These two components can be fit by two
exponential functions, and the value of $P$ is conservatively chosen so that
coincident events due to noise are less than 1\% of all triggers. For
regular camera operation, $P$ is 5.5~p.e., which ensures stable operation even
at higher levels of NSB light, up to $\sim 250$~MHz photon rate. Note that at
this value of $P$, the single-telescope L2 trigger rates are already in the
noise, with rates well in excess of 1 kHz, which with the old cameras would
have caused more than 36\% of the events to be lost due to dead-time, a figure
that becomes around $\sim1\%$ with the new cameras, thanks to the new
\nectar-based readout.

At the array level, it is important to measure the signal round-trip time
between the central trigger and the camera, in order to adjust the fixed part of the
central trigger coincidence delays, which also vary depending on the pointing direction. 
This is done by sending a trigger signal via
optical fiber from the central trigger to the DIB, which then replies to it.
The difference between the time of sending and that of receiving is measured at
the central trigger with an oscilloscope. On average, the round-trip time was
reduced by $\sim 300$ ns with respect to the original cameras.


\section{Performance}
\label{sec:perf}
We report some of the most significant performance metrics for the new Cherenkov
cameras in this section. Some of them were measured in the lab, prior to the
installation of the cameras, others in the field in Namibia, during or after
commissioning.

Efforts are ongoing to fully characterize the performance of the new cameras in terms of gamma-ray sensitivity using simulations and standard
candle data; and to exploit the several new features they offer. The results will be made available in upcoming publications by the H.E.S.S.~collaboration.

\subsection{Analogue front-end}
\label{sec:perf_analogue}
\begin{figure}
  \centering
  \includegraphics[width=0.45\textwidth]{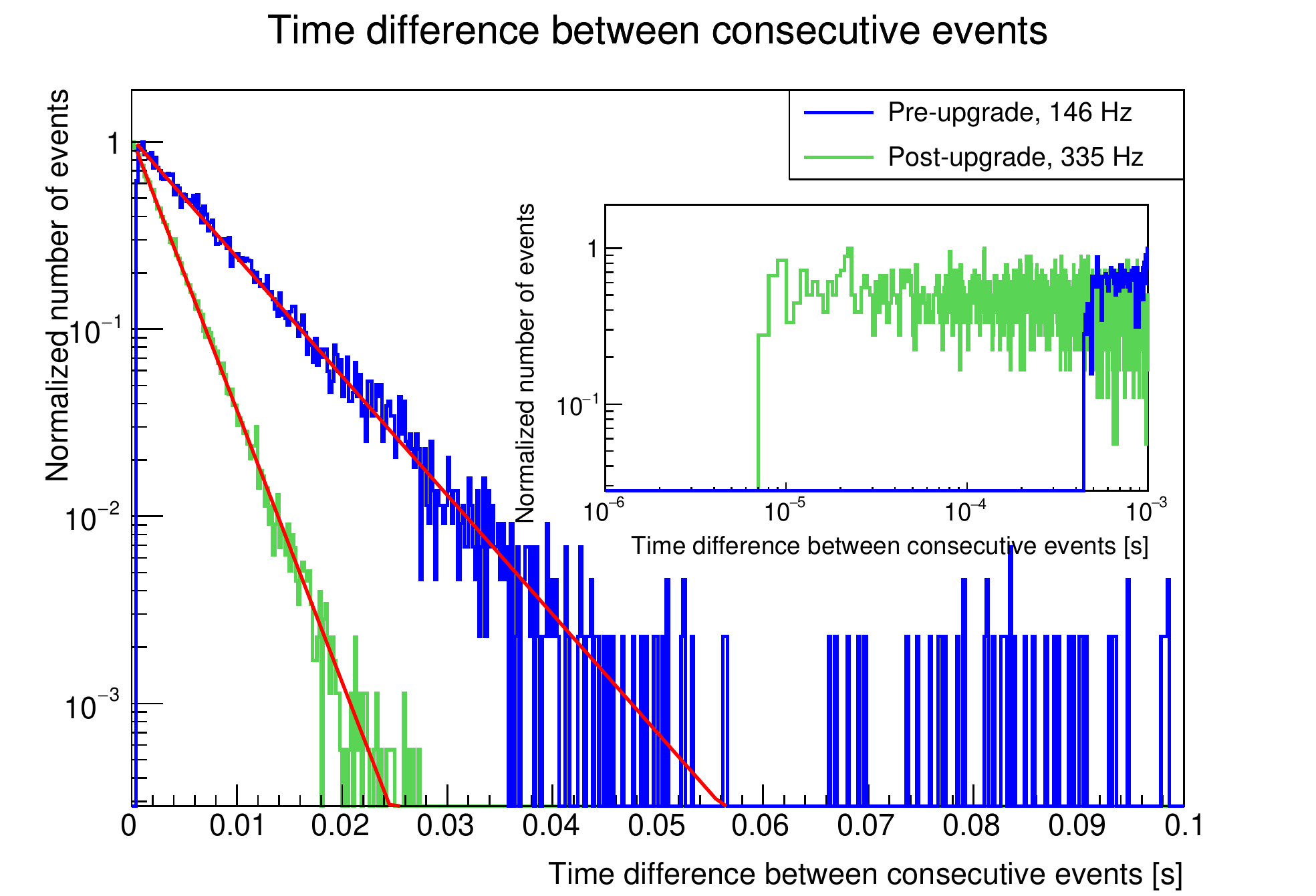}
  \includegraphics[width=0.45\textwidth]{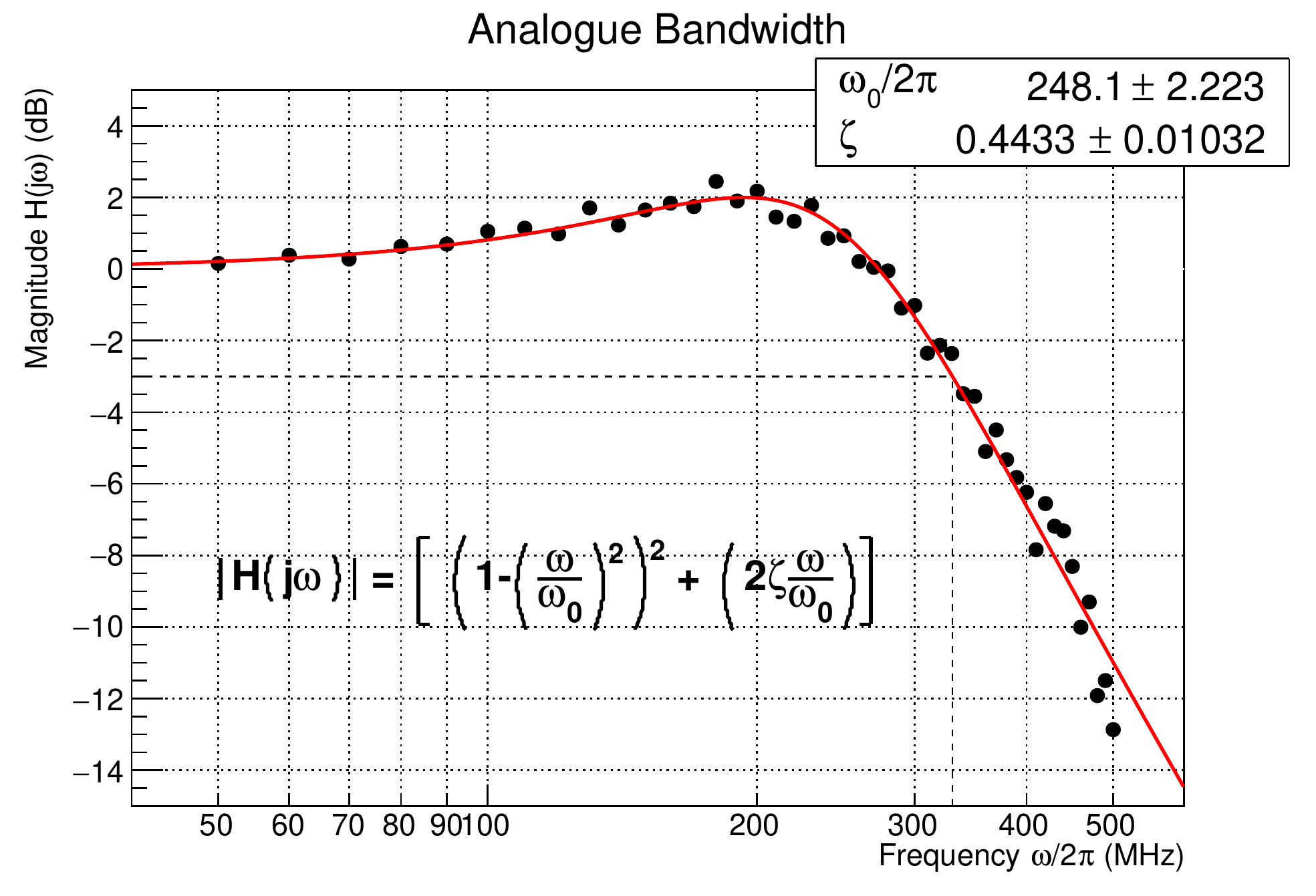}
  \caption{Left: Normalized distributions of the time delay between two
  consecutive events in the original (blue) and upgraded (green) \hessI \
  cameras. The two runs from which this plot is taken had different average trigger rates.
  The log-log plot in the inset shows a zoom of the first millisecond with finer
  binning.  From this plot the dead-time can be estimated as the shortest time
  between two consecutive events, i.e. the position of the left-most bin with
  entries. Before the upgrade it was $\sim450$~{\textmu}s, after the upgrade
  $\sim7$~{\textmu}s. Right: measurement of the end-to-end Bode magnitude plot of
  the readout electronics, The fit function form employed
  $\left|H\left(j\omega\right)\right|$ is the modulus of the transfer function
  of a second-order underdamped system $H(s) = \omega_0^2 / (s^2 + 2 \zeta
  \omega_0 s + \omega_0^2)$, with damping ratio $\zeta$ and eigenfrequency
  $\omega_0$. The dashed lines show the -3~dB point of the plot, corresponding
  to $\omega/2\pi \sim 330$~MHz. The measurement was done injecting a pure sine wave
  of varying frequency into the system and measuring the relative amplitude of
  the digitized waveforms.}
    \label{fig:deadtime}
\end{figure}

The dead-time of a \nectarchip\ when reading the nominal 16 cells region of
interest is about $1.6$~{\textmu}s
\cite{delagnes_nectar0_2011,nectardatasheet}. 
However, the minimum safe time interval between two events is greater than the
nominal dead-time of the \nectarchip,  because of the trigger signal generation
and the chip readout process on the FPGA take $\sim4$~{\textmu}s. Also, for the
version of the \nectarchip\ used here, the first 16 readout cells have to be read out and
discarded because of stale values, adding another $1.6$~{\textmu}s to the
dead-time. Due to all this, the hold-off time is set to: $t_{b} = 4 +
(n+n/16)\times0.1$~ns, where $n$ is the total number of \nectar\ cells read out
($n=32$ for regular observations). This can be appreciated in
\fref{fig:deadtime}, left, which shows that the overall dead-time of the
upgraded \hessI\ cameras, measured from the distribution of the time difference
of two consecutive events during a regular observation run is
$\sim7.2$~{\textmu}s. 

The nominal analogue bandwidth of the \nectarchip\ is 410~MHz
\cite{delagnes_nectar0_2011,nectardatasheet}. The design of the analogue
electronics uses components matching or exceeding that bandwidth. The end-to-end
-3~dB bandwidth of the readout is $\sim 330$~MHz, more than four times higher
than in the previous camera, see \fref{fig:deadtime}, right. One can see the
benefit of such a high bandwidth in the sampled PMT pulse shape shown in the
left panel of \fref{fig:gain_calib}, where the FWHM is less than 3.5~ns. Such
narrow peaks allow a better determination of shower time profiles, which can
be used to improve the sensitivity of the analysis \cite{ALIU2009293,STAMATESCU2011886}. 

The design of the analogue part of the readout was optimized for low noise.
The pedestal noise, which is the RMS of the value of a single \nectar\ cell in the
presence of no input signal is on average $\sim4$~ADC counts, or $\sim2$~mV. This
corresponds to the pure electronic noise of the front-end. For the typical PMT gain of $2.72\times10^5$, and charge integration window of 16 samples, the electronic noise of front-end and PMT combined has on average a RMS of $\sim16$~ADC counts, or $\sim0.2$~photo-electrons (p.e.). 
 This is measured routinely during single photo-electron calibration
runs, as the RMS value of the pedestal distribution (see \fref{fig:gain_calib},
right). The noise intensity varies with the
position of the channels inside the drawer: the channels whose amplifiers are
physically located in the front of the drawer, closer to the PMTs (channels 4--7
and 12--15) have higher electronic noise ($\sim0.25$~p.e. RMS) than the ones who
are in the back of the drawer ($\sim0.15$~p.e. RMS), forming two distinct
populations. This is likely due to different noise pick-up along the routes of the
traces on the analog circuit board. It is anyway not a problem because at the chosen gain, the single-electron signal is distinguishable in any case, being always at least 3 times higher than the noise.

\begin{figure}
  \centering
  \includegraphics[width=\textwidth]{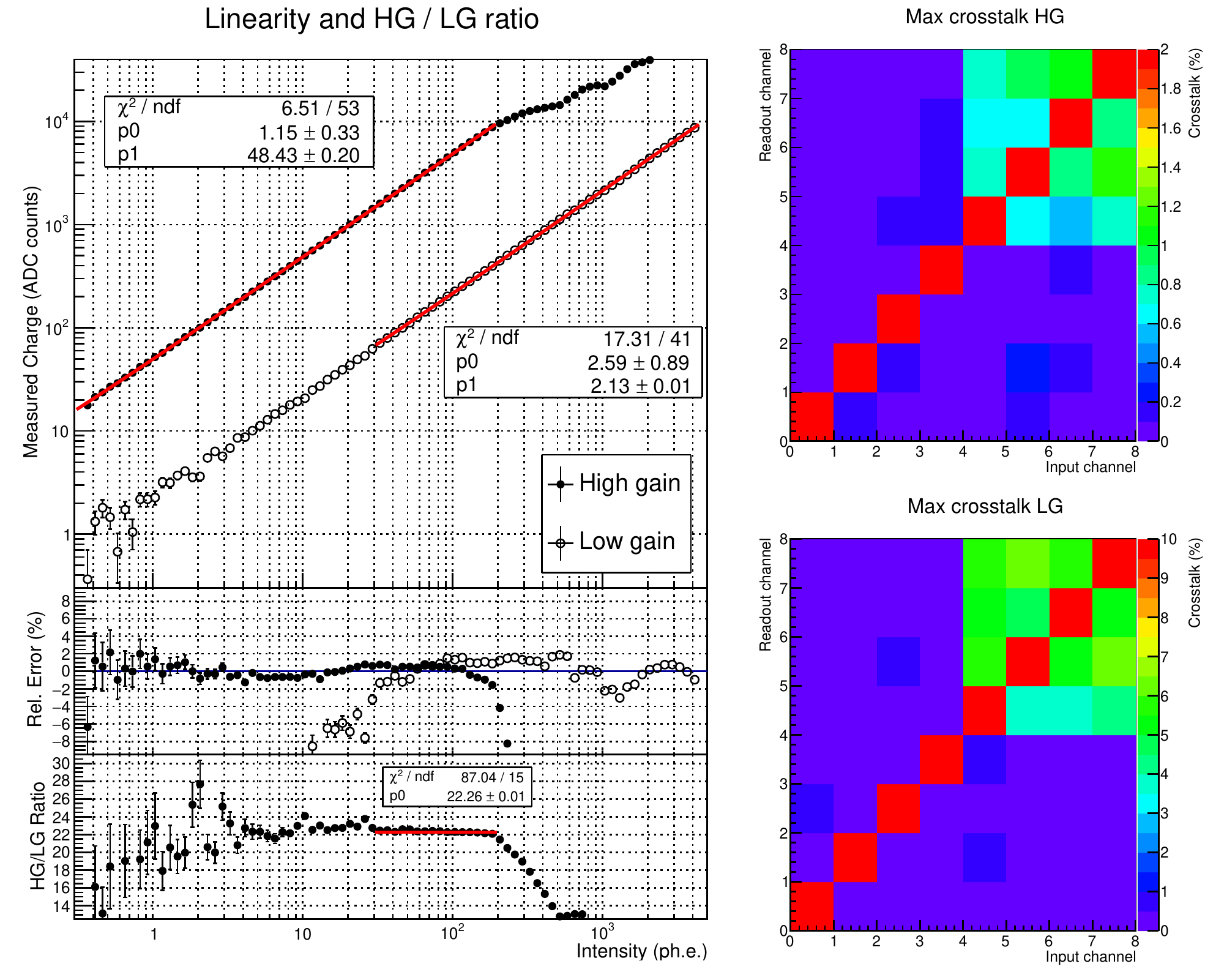}
  \caption{Left: Linearity of a typical readout channel. The top frame shows
    the recorded charge versus the input pulse intensity, both for high (black circles)
    and low (white circles) gain. Two linear functions 
    (red lines) fit this data, their fit parameters and $\chi^2$ values are displayed next to them. The fit residuals are displayed in the middle panel. The bottom panel shows
    the ratio between the two gains, and a fit to a constant value in the overlapping range 30--200~p.e..  Right: Maximum cross-talk inside one analogue
    board, for both high (top) and low (bottom) gain. The cross-talk is computed as the ratio $C(i,r)$ between charges recorded in any pair of channels; 
    the x-axis corresponds to the channel $i$, where the test pulse is applied, the y-axis to the readout empty channel $r$.}
    \label{fig:linearity}
\end{figure}

The linearity and cross-talk of the readout were measured by recording a
pre-calibrated, PMT-like pulse of variable intensity. The results, which can be
seen in the left panel of \fref{fig:linearity}, show that non-linearities in
both high gain and low gain amount to less than 2\%. The linear range of the
high gain is 0.3--200~p.e. and that of the low gain is 30--4,200~p.e.: the total
readout dynamic range is greater than 80~dB. The ratio between high and low gain is
$\sim22$ between 30 and 200~p.e. (see \fref{fig:linearity}, left, bottom panel).

The data mentioned above was also used to characterize the cross-talk between
two channels on the same analogue board. For the high gain channel the cross-talk is typically less than 0.5\%, and never larger than 1\%; for the low gain is at most 7\%
(see \fref{fig:linearity}, right).  It is measured using the largest PMT-like pulse 
inside the linear range of each
gain, and taking the ratio $C(i,r) = Q_{r}/Q_{i}$ between the charge recorded in 
an empty channel ($Q_{r}$) to that measured in the input channel ($Q_{i}$). 
Similarly to the electronic noise, the cross-talk is also larger
for the front channels (4--7 and 12--15) than for the back ones. This problem was
studied in detail with auto-correlation and frequency domain analysis of the recorded noise, but no obvious cause was found \cite{shiningayamwe_investigating_2017}.

\subsection{Trigger}

The optimization of the trigger described in the previous section increased the
fraction of events triggered stereoscopically with CT5 by more than a factor of
two. Before the upgrade it was 20\%, after the upgrade it is 44\%. This is a
direct consequence of the reduced deadtime of the camera due to the \nectarchip,
which allows the camera pixel threshold to be lowered substantially.

In the case of observations with a low NSB light intensity in
the field of view (i.e. an average pixel photon rate across the camera of less
than 100~MHz), the nominal pixel trigger threshold can be lowered by 1~p.e, to
4.5~p.e. Preliminary studies on simulations showed that this simple adjustment
results in marginal improvements in terms of threshold trigger effective area,
which were not deemed sufficient to justify the manpower investment in the
production and maintenance of a full new set of simulations and instrument
response functions.

The next-neighbour alternative trigger architecture was also tested and
simulated, but it was found not to deliver a substantially improved performance
with respect to the default 3-majority scheme. The performance of the pseudo-sum
trigger alternative is still under study due to the higher number of parameters
to optimize and difficulty of implementing a realistic simulation.

\subsection{Readout and slow control}
\label{ssec:readout_slow}
The \nectarchip\ design, the modularity of the camera, the advanced driver for
the FPGA--ARM memory bus exploiting direct memory access (DMA) technology and
the ample software buffering allow for a maximum achievable data acquisition
rate with default settings (i.e. readout and storage of integral charge and
timing information) of around 10~kHz per telescope. This is about twenty times higher than
the usual CT1--4 acquisition rate during regular observations. It was
determined by field tests under realistic conditions. \\ The bottleneck is the
transfer of data to the \hess \ main DAQ program, because the network bandwidth
is only 1~Gb/s. Performance tests on a 10~Gb/s network showed that the cameras
could sustain a constant individual data acquisition rate in excess of 50~kHz. The system can sustain short bursts of events at a much higher rate by buffering the data in the RAM of the ARM computer and of the camera server. This can be very important for some
physics cases, such as transient events and especially GRBs.
%

The improvement of the new camera readout system allows to configure the readout
so that full waveforms of up to 48~samples are stored alongside the integrated
charge over 16~ns and the timing information. This is expected to be beneficial
in the reconstruction of inclined or large impact parameter showers with
energies larger than 1~TeV, for which the arrival time dispersion of Cherenkov
light at the telescope is greater than 16~ns. This readout mode increases the
amount of transmitted data by a factor $\sim17$ (each drawer sends 51 data
blocks instead of the usual 3). In order to keep up with the usual data
acquisition rates (up to 700~Hz) when using this acquisition mode, the
additional waveform data must be stored on the camera server hard disks, and is
transmitted to the \hess \ DAQ off-line on the following day. This mode is only
used for selected targets, due to the much greater amount of data created when
it is active. Initial results on the performance of this readout mode are reported in \cite{zorn_sensitivity_2019}.

Regarding the slow control software performance, stress tests performed on the
Apache Thrift RPC framework operating in the busy DESY lab network showed that
it is capable of sustaining rates of 10,000 single point-to-point request/replies
per second for more than 12 hours with no failures. One-to-many requests, such
as distributing a command or collecting information from all drawers, are
handled on the camera server by spawning one thread per connection. This
strategy allows for a command distribution latency of $\sim8$~ms.

\subsection{Commissioning and long-term stability}
\begin{figure}
  \centering
  \includegraphics[height=.32\textheight]{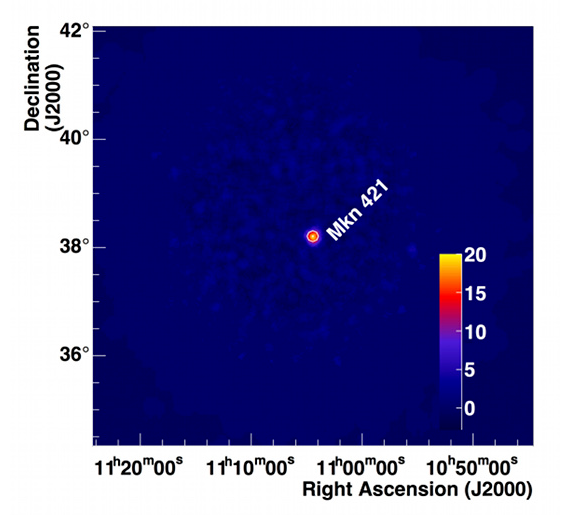}
  \includegraphics[height=.30\textheight]{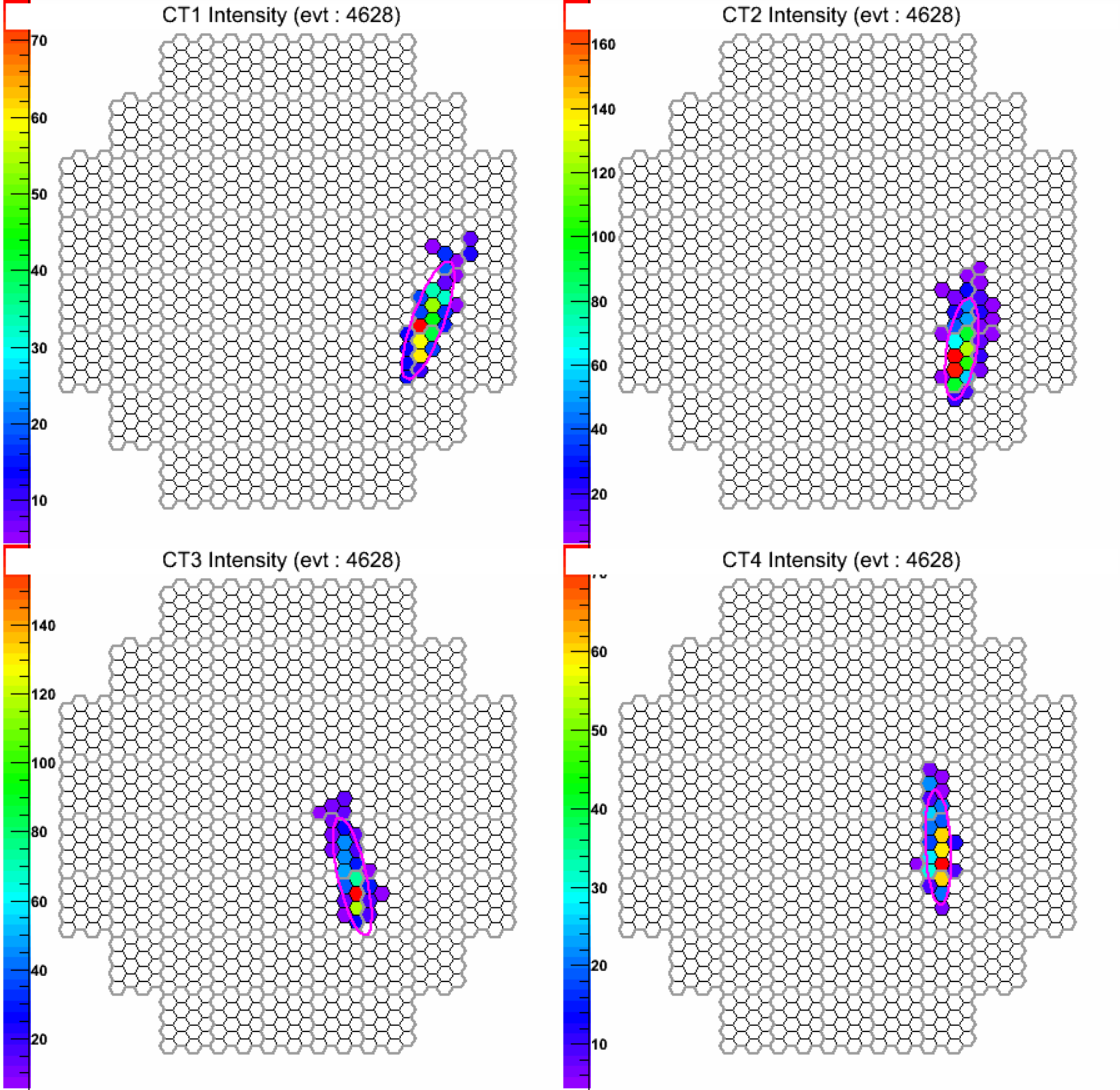}
  \caption{Left: Significance sky map of Mkn~421, a well-known TeV gamma-ray
    emitting blazar, observed during the commissioning of the \hess~upgrade
    ~cameras.  Right: Example 4-telescope event recorded with the upgrade
    cameras. Figure adopted from \cite{SOM_Mkn421}.}
  \label{fig:SOM_Mkn421}
\end{figure}

The upgrade of the first camera, that of CT1, was carried out in July/August
2015. This was followed by an extended integration and commissioning period of
9~months. During this period of time, many bugs and problems were ironed out,
while the rest of the array (CT2-5) continued scientific observations with
minimally degraded performance. This strategy allowed us to compare old and new
cameras after the first one was completely commissioned. The other three
upgraded cameras were installed in September/October 2016 and underwent a much
shorter commissioning phase of four months. In January 2017, a bright flare from
the well-known Mkn~421 blazar was observed by \hess~ using the new upgraded
cameras, following an alert reported by the HAWC collaboration \cite{2017ATel.9946....1M}. 
About 2 hours of data were collected during this observation. The
preliminary processing of the data using two independent analysis pipelines
revealed a clear detection with a significance of 16$\sigma$. This was the first
detection of a TeV gamma-ray source using the \nectarchip\ technology
\cite{SOM_Mkn421} (see \fref{fig:SOM_Mkn421}, for a significance sky map of this
detection and an example event).  The upgraded cameras have been employed in
routine observations since January~2017, and since then have achieved an average 
weather-corrected data taking efficiency of 98.5\%.


\section{Conclusion}
\label{sec:conclusions}
The four upgraded cameras of the 12-meter \hess\ Cherenkov telescopes were
successfully deployed on site in 2015 and 2016. They are equipped with a new
\nectar-based readout technology that substantially reduces the dead-time by a factor of 60 from $\sim450$~{\textmu}s in the previous system to $\sim7$~{\textmu}s in the new cameras.  Furthermore, the new design allows for a more robust, versatile and efficient operation and maintenance, leading to improved performance and reliability. All components of the cameras were tested, integrated and calibrated, and their performance was validated in the field. The camera configuration was optimized, resulting in more than twice the amount of stereoscopically recorded showers by the \hess\ array. 

The achieved average data taking efficiency of the cameras is 98.5\%. No major problems due to ageing were found during an ordinary maintenance campaign that took place in early 2019. 
Thus, all the primary goals of the project have been achieved.

In addition, the new cameras offer the possibility of using more sophisticated and flexible trigger 
and readout algorithms. The most promising of these new possibilities is to record fully sampled
waveforms, which is being explored intensively in current observation campaigns and will be
reported on in the future.

The new cameras are foreseen to be in use in the \hess\ experiment for its remaining lifetime.

\section*{References}

\bibliographystyle{model1-num-names}
\bibliography{sample}

\section*{Acknowledgements}      
The support of the Namibian authorities and of the University of Namibia in
facilitating the construction and operation of H.E.S.S. is gratefully
acknowledged, as is the support by the German Ministry for Education and
Research (BMBF), the Max Planck Society, the German Research Foundation (DFG),
the French Ministry for Research, the CNRS-IN2P3 and the Astroparticle
Interdisciplinary Programme of the CNRS, the U.K. Science and Technology
Facilities Council (STFC), the IPNP of the Charles University, the Czech
Science Foundation, the Polish Ministry of Science and Higher Education, the
South African Department of Science and Technology and National Research
Foundation, and by the University of Namibia. We appreciate the excellent work
of the technical support staff in Zeuthen, Durham, Hamburg, Heidelberg,
Palaiseau, Paris, Saclay, and in Namibia in the construction and operation of
the equipment.

The authors would also like to thank the anonymous reviewers for their valuable
insights.

\end{document}